\documentclass{article}
\usepackage{macros,soul}
\date{\today}
\author{Mayank Raj, Lianghao Cao, Andrew Stuart, Kaushik Bhattacharya}
\affil{Division of Engineering and Applied Sciences, California Institute of Technology, Pasadena, CA 91125, USA}

\title{A Neural-Network Framework to Learn History-Dependent Constitutive Laws and Identifiability of Internal Variables} 
\begin{document}
\begingroup
\renewcommand{\thefootnote}{}
\footnotetext{\texttt{
mraj@caltech.edu (Corresponding author),
lianghao@caltech.edu,
astuart@caltech.edu,
bhatta@caltech.edu}}
\endgroup
\numberwithin{equation}{section}
\numberwithin{figure}{section}
\numberwithin{theorem}{section}
\mathtoolsset{showonlyrefs}
\maketitle
\begin{abstract}
The identification of constitutive laws is ubiquitous in engineering: in modeling of materials where experimental data are fitted to mathematical models or learning surrogate models to beat the FE\textsuperscript{2} computational cost of multiscale numerical simulations. However, these models of constitutive laws, unless equipped with a potential formulation, are not necessarily consistent with (a) the second law of thermodynamics; (b) stability of the material under extreme applied strain; and (c) the mathematical theory underpinning the existence of solutions of the governing equation.
In this work, we present a causal and energetic formulation, consistent with aforementioned properties, of learning a history-dependent constitutive law. This characterization of the class of internal variables sheds light on the
equivalence class of equivalent surrogate models for the constitutive law. We show that the internal variables that are learned from the data are unique up to a linear transform.  The framework is deployed to learn the Taylor-averaged response of a polycrystalline magnesium unit cell. We achieve 2\% relative error in the prediction of the Taylor-averaged response. 
\end{abstract}

\section{Introduction}
\label{sec:Intro}
Modeling the mechanical response of a material at the length scale relevant to its engineering application, known as the macro-scale, is often complicated. This is because the macro-scale response of a material is a cumulative result of physical processes occurring at a length scale finer than the scale of engineering applications. Moreover, the physics of the mechanical response of a material is often simpler and, hence, much better understood at a fine length scale. In this context, homogenization refers to obtaining the effective macroscopic response of a material based on knowledge of its microscopic behavior. Although the analytic derivation of equations governing the macroscopic response from known microscopic physics may be possible in a simplistic setting \citep{bensoussan2011asymptotic,pavliotis2008multiscale,KOVACHKI2022104156}, doing so for many classes of materials is often intractable . Therefore, a computational approach to homogenization, where data from a fine-scale process may be used to model the constitutive law at a coarse scale, is indispensable. In the recent years, data-driven methods have been explored to beat the FE\textsuperscript{2} computational cost of two-scale homogenization. Applications of this idea can be seen in the \cite{ASAD2026118666} for a woven fabric of viscoelastic material, and \cite{LIU2023105329} for plastic material with history-dependent stress-strain response. In the data-driven approach to homogenization, the response of the unit-cell is fit to a neural network model that is used as constitutive law in a macroscale FE solver.

Apart from homogenization setting, neural networks are also being used to fit a constitutive model to experimental data. Historically, many finite-deformation constitutive laws have been developed to model the macroscopic non-linear response of materials. However, these classical non-linear constitutive laws are designed for very specific materials and usually do not generalize well to other similar, but different, materials. 
Modeling finite-deformation constitutive laws become even more complicated for material behavior that depends on the history of the applied load. History-dependent behavior may arise in cases such as viscoelasticity \citep{Francfort1986}, plasticity \citep{Simo_hughes}, and in the homogenized response of rapidly varying continua \citep{BRENNER20131824}. Hence, there has been an attempt to leverage generalization ability of neural networks to learn the constitutive response of a wide class of materials.

In this work, the constitutive law of the homogenized response of a polycrystalline unit cell, subject to finite deformation, is learnt using NN models. Mechanical response of metals under finite deformation is inelastic and hence is modeled by introducing a set of internal variables. Internal variables are used in thermodynamics formalism of mechanics for modelling inelastic materials. The idea is to expand the set of thermodynamic state variables by a set of internal variables $(\xi_1, \dots, \xi_n)$ to model the transition between two thermodynamic equilibrium states. The first application of internal variables in mechanics can be traced back to \cite{Coleman}, where thermodynamics restriction on the internal variables is derived  and its connection with dynamical stability is analyzed. Application of this theory in particular in metal plasticity is done in \cite{RICE1971433}, where internal variables are used to represent the average behavior of local microstructural rearrangements due to dislocations. A historical review of internal state variable theory can be found in \cite{HORSTEMEYER20101310}. In the classical approach, internal variables and their evolution equation are postulated apripori as a part of constitutive law. Their evolution equation is calibrated based on experimental observations such as yield in metal plasticity. At the same time, internal variables could also arise out of an approximation theoretic approach towards approximation of integrals that capture history-dependence. The physical meaning of internal variables depends on their application. In metal plasticity, internal variables are postulated to represent plastic strain that in-turn is believed to capture average deformation due to dislocation rearrangements. However, in the case of viscoelasticity, internal variables arise out of approximation of integral \cite{Francfort1986,Margaret2023} and hence, do not represent any physical or experimentally measurable quantity. In the more recent data driven approach towards constitutive modeling, internal variables and their evolution law are discovered from data.

Recurrent neural network based constitutive laws that directly map strain to stress or displacement to force do not impose a priori by architecture (a) the second law of thermodynamics, (b) material stability, and (c)  mathematical conditions that guaranty existence of solutions of the governing equations. These mathematical conditions are described for hyperelastic material in \citep{ball1977nonlinear, Ball1976} and extended to history-dependent material in \citep{Mielke2003,ORTIZ1999419}.   The existence of a solution for history-dependent governing equations relies on the polyconvexity of the energy potential \cite{Ball1976} and the coercivity of both the energy and dissipation potentials \citep{Mielke2004,FrancfortMielke2006,Mainik2005,Mielke2018}.  The research reported in this paper builds on the existing literature on modeling 
the homogenized macroscopic mechanical response of a polycrystal. The response is 
assumed to be viscoplastic, and a mapping from strain history to stress is
sought.  To this end, a novel NN framework, based on internal variable theory, 
is introduced. The main contributions of the work are now summarized.
\begin{enumerate}[label=(C\arabic*)]
    \item A NN framework is introduced, consistent with the second law of thermodynamics, material stability, objectivity and the theory underpinning existence of a solution of the non-linear momentum balance equation, to learn a constitutive law mapping strain history to stress.

    There have been recent work where some of these constraints are imposed: polyconvexity, material stability, and objectivity for hyperelastic material in \cite{Asad2022,KLEIN2022104703}, the second law of of thermodyamics in the case of one-dimensional elasto-plastic material in \cite{FLASCHEL2025} and two-dimensional viscoelastic material in \cite{ASAD2023116463}. However, this framework has not been tested on a three-dimensional metal visco-plasticity in two-scale homogenization setting. Latest approach towards visoplastic homogenization is introduced in \citet{LIU2023105329,Karimi2024} where NN framework invariant in time-discretization is  to learn history-dependent constitutive laws by learning evolution of internal variables. However, the architecture there is not consistent with the foregoing mathematical properties. 

    \item The polyconvex model is trained on the space of symmetric deformation gradient and is extended to the space of non-symmetric deformation gradient in a manner consistent with objectivity.

    The current literature shows there are not abstract forms of constitutive laws that are both polyconvex and objective. The both are simultaneoulsy achieved by invariant-based models where more assumptions on material symmetry are made such as in \cite{Schroder2010}. Data-driven approach towards imposing both polyconvexity and objectivity uses an objectivity-based  additional term in the loss function as in \cite{KLEIN2022104703}. The data-driven approach performs poorly in terms of stress-prediction \cite{KLEIN2022104703}, hence we present an approach where objectivity is achieved by definition
of energy function along with polyconvexity.
    \item The framework is demonstrated by fitting the Taylor-averaged response of a polycrystalline magnesium microstructure, achieving a relative error of 2\%, resulting in a substantial speed-up over the original model, and without any compromise on accurate prediction.
    \item Empirically, it is shown that the internal variables are identifiable
 up to a linear transform; theoretical results supporting this observation are
 provided.
 
    In the case of inelastic material, it has been reported that internal variables learned from data are not unique \cite{FLASCHEL2025};  however, there is no theoretical charecterization of the set of these different but equivalent set of internal variables.
 up to a linear transform; theoretical results supporting this observation are
 provided.
\end{enumerate}
 The paper is organized as follows. Section~\ref{sec:background} provides the mathematical background concerning modeling inelastic history-dependent mechanical response, which includes the second law of thermodynamics,
polyconvexity, growth properties of potential pertinent to material stability, and objectivity. We also present, in Section~\ref{sec:background}, the application of the Legendre transform in the formulation of explicit kinetics. 
Section~\ref{sec:M} presents the NN framework, (C1), that covers a description of the NN architecture used to parameterize various potentials, methods to enforce objectivity and polyconvexity of the energy potential, the method to enforce convexity 
of the dissipation potential, and the functions used to enforce the growth of potentials; in the same section we also provide details of data generation and training (C2). Section~\ref{sec:rd} presents (C3) and (C4), demonstrating the advantage of the proposed energy decomposition and efficacy of the
resulting framework in terms of achieved prediction accuracy. The empirical verification of the identifiability of internal variables up to linear transform and supporting theoretical results (C5) are also contained in Section~\ref{sec:rd}. Finally, a concluding discussion on limitations and further scope of this work is presented in Section~\ref{sec:conclusion}. 
 
\section{Mathematical Framework}
\label{sec:background}
Let $\tensor{F}(t) \in \Rdd$ and $\tensor{P}^\dagger(t) \in \Rdd$ denote,  
respectively, the deformation gradient and first Piola-Kirchhoff stress defined
by the material of interest. The behavior of a viscoplastic 
material is non-Markovian map of the form:
\begin{align}
    \Psi^{\dagger}: \cbkt{ \tensor{F}(\tau): \tau \in \bkt{0, t}} \mapsto \tensor{P}^\dagger(t), \label{eq:1}
\end{align}
The superscript $\dagger$ is used throughout this article to distinguish the 
true map, from its approximation.  In Section~\ref{ssec:MCL}, we lay out a Markovian formulation to model the history-dependent constitutive law, and state the physical properties desired from such a constitutive law. An explicit formulation of the dynamics, capturing the history-dependence and employing
internal variables, is presented in Section~\ref{ssec:Leg}.


\subsection{Markovian Constitutive Laws}\label{ssec:MCL}

The map \eqref{eq:1} is non-Markovian when used as a constitutive law in the balance of linear momentum equation. However, for interpretability and computational efficiency, it is desirable to deploy Markovian models whenever possible. Of course, the choice of such a Markovian approximation must allow for an adequate 
representation of physics. Markovian approximation of non-Markovian models is often done by introducing a set of internal variables $\xi(t) \in \R^{n}$. 

In an energetic formulation, the stress and evolution of internal variables are determined from derivatives of the energy 
density potential $\cW: \Rdd \times \R^{n} \to \R$ and the 
dissipation potential $\widehat{\cD}: \R^{n} \to \R$. Specifically 
\begin{subequations}
\label{eq:3}
    \begin{align}
        \tensor{P}(t) &= \dfrac{\partial \cW(\tensor{F}(t), \xi(t))}{\partial \tensor{F}}, \label{eq:3a}\\
        \dfrac{\partial \widehat{\cD}{(\dot{\xi}}(t))}{\partial \dot{\xi}} &= -\dfrac{\partial \cW(\tensor{F}(t), \xi(t))}{\partial \xi},\quad \xi(0) = 0. \label{eq:3b}
    \end{align}
\end{subequations}
We use $\dot{()}$ to denote the time derivative, as in $\dot{\xi}(t)$. 
It should be noted that $\tensor{F}, \xi, \dot{\xi}$ are used as dummy variables for arguments of $\cW$ and $\widehat{\cD}$; whereas $\tensor{F}(t), \xi(t)$ and $\dot{\xi}(t)$ represents these variables at time $t$. We follow this convention throughout this paper. This energetic  Markovian formulation leads to several desirable properties that we now
list.

\textbf{The Second Law of Thermodynamics.}
The second law of thermodynamics, in the form of the Clausius--Duhem inequality, states that the dynamics of the process must be dissipative. Assuming the process to be isothermal, the Clausius--Duhem inequality can be written as 
\begin{align}
     \dfrac{\partial \widehat{\cD}{(\dot{\xi}}(t))}{\partial \dot{\xi}} \cdot \dot{\xi}(t) \geq 0. \label{eq:6}
\end{align}
A sufficient condition that ensures that \eqref{eq:6} is satisfied for all $t \in \R^+$ is that $\widehat{\cD}$ be a convex function with a minimum at zero. Method to impose convexity is presented in Section~\ref{ssec:NNA}.

\textbf{Objectivity.} The mechanical response of a material must be indifferent to the reference frame of an observer. As a consequence, it must satisfy $\cW\bkt{\tensor{QF}, \cdot} = \cW\bkt{\tensor{F}, \cdot}$ for all $\tensor{Q} \in \mathrm{SO3}$. Therefore, $\cW$ must admit a representation such that $\cW(\tensor{F}, \xi) = \widehat{\cW}(\tensor{C}, \xi)$ where $\tensor{C} := \tensor{F^{\top}F}$, is called the Cauchy strain tensor. Equations \eqref{eq:3a} and \eqref{eq:3b} can be written in terms of the Cauchy strain tensor to satisfy objectivity. To this end, we have 
\begin{subequations}
\label{eq:ds1}
    \begin{align}
        \tensor{S}(t) &= \dfrac{\partial \cW(\tensor{C}(t), \xi)}{\partial \tensor{C}} \label{eq:10a}\\
        \dfrac{\partial \widehat{\cD}(\dot{\xi}(t))}{\partial \dot{\xi}} &= -  \dfrac{\partial \cW(\tensor{C}(t), \xi(t))}{\partial \xi},\quad \xi(0) = 0, \label{eq:10b}
    \end{align}
\end{subequations}
where the symmetric tensor $\tensor{S} \coloneqq \tensor{F}^{-1}\tensor{P}$ in \eqref{eq:10a} is called the second Piola-Kirchhoff stress. The benefit of formulating the problem in terms of $\tensor{C}$ is that it reduces the dimensionality of the computation, as both $\tensor{S}$ and $\tensor{C}$ are symmetric (in contrast to their counterparts $\tensor{P}$ and $\tensor{F}$). At the same time, a drawback of this formulation is that, without making any further simplifying assumptions, it is hard to restrict $\cW$ to be a polyconvex function of $\tensor{F}$. 

\textbf{Polyconvexity.}
The notion of polyconvexity was introduced in the seminal work on existence theorems for nonlinear elasticity momentum balance PDEs \citep{Ball1976, ball1977nonlinear}. Energy density $\cW(\tensor{F}, \xi)$ is polyconvex in $\tensor{F}$ iff it admits a representation  $\cW(\tensor{F}, \xi) = g(\tensor{F}, \mathrm{adj}\, \tensor{F}, \det \tensor{F}, \xi)$ where $g(\cdot, \cdot, \cdot, \xi)$ is a convex function for all $\xi \in \R^{n}$. The polyconvexity of the energy density potential allows for 
characterization of a wide range of physical assumptions under which the non-linear momentum balance equation has a solution. Polyconvexity is imposed by imposing convexity, which is discussed in Section~\ref{ssec:NNA}, and more discussion on polyconvex model is done in Section~\ref{ssec:PNN}. Apart from polyconvexity, the existence of a solution also relies on certain growth properties of the potential, which are discussed next.

\textbf{Material Stability.} Apart from being a necessity for the existence theorems as mentioned in the previous paragraph, other growth conditions may also be desired to prevent unrealistic mechanical behavior when a material is subjected to extreme strain; hence the name `material stability'. Here, we discuss two different growth conditions that have physical interpretations and significance. It is expected that 
\begin{align}
    \dfrac{\cW(\tensor{F}, \xi)}{\abs{\tensor{F}}^3} \to \infty  \quad \text{as} \quad \abs{\tensor{F}} \to \infty; \label{eq:7}
\end{align}
where $\abs{\cdot}$ represents the Frobenius norm. The limit in \eqref{eq:7} implies that a line segment of positive length cannot be produced from an infinitesimal cube using a finite amount of energy. It is to be noted that \eqref{eq:7}, together with objectivity, translates to 
\begin{align}
    \dfrac{\widehat{\cW}(\tensor{C}, \xi)}{\abs{\tensor{C}}^{3/2}} \to \infty  \quad \text{as} \quad \abs{\tensor{C}} \to \infty.
\end{align}
Another growth condition, which is not necessary for the existence theorems,
but is physically desirable, is as follows
\begin{align}
    \widehat{\cW}(\tensor{C}, \xi) \to \infty \quad \text{as} \quad \det \tensor{C} \to 0^{+}. \label{eq:9}
\end{align}
The limit in \eqref{eq:9} ensures that a piece of material cannot be 
compressed to a point with a finite amount of energy or compressive stress. These limiting behaviors are imposed using certain functions that are presented in Section~\ref{ssec:GF}.

  We constrain both these models to be consistent with the second law of thermodynamics and material stability. Before we describe the NN parameterization of $\cW$, $\cW$ ,and $\widehat{\cD}$ and the training procedure, it is worthwhile to note that both \eqref{eq:3b} and \eqref{eq:10b} are implicit ODEs. However, owing the  convexity of $\widehat{\cD}$ as a consequence of the second law of thermodynamics, it is possible to derive an equivalent explicit form of $\eqref{eq:3b}$ and $\eqref{eq:10b}$. This is the subject of discussion in the next Section~\ref{ssec:Leg}.

\subsection{Legendre Transform and Explicit Formulation of Dynamics}\label{ssec:Leg}
Since $\widehat{\mathcal D}$ is convex, we use convex duality to (2.7b); the same argument holds for (2.3b). To this end, we work with the Legendre transform
\citep{simo1998computational} of $\widehat{\cD}$: we define $\cD : \R^{n} \to \R$ by
\begin{align}
    \cD(d) = \sup_{\dot{\xi} \in \R^{n}} \bkt{\dot{\xi} \cdot d - \widehat{\cD}(\dot{\xi})}.
\end{align}
If $\widehat{\cD}$ is continuously differentiable,
\begin{align}
    d = \dfrac{\partial \widehat{\cD}(\dot{\xi})}{\partial \dot{\xi}};\quad \dot{\xi} = \dfrac{\partial \cD(d)}{\partial d}. \label{eq:13}
\end{align}
Using \eqref{eq:13},  \eqref{eq:3} can be written as 
\begin{subequations}
\begin{align}
          \tensor{P}(t) &= \dfrac{\partial \cW(\tensor{F}(t), \xi(t))}{\partial \tensor{F}} \label{eq:14a}\noeqref{eq:14a}\\
        d(t) &= -  \dfrac{\partial \cW(\tensor{F}(t), \xi(t))}{\partial \xi}\label{eq:14b}\noeqref{eq:14b}\\
        \dot{\xi}(t) &= \dfrac{\partial \cD(d(t))}{\partial d},\quad \xi(0) = 0. \label{eq:14c} \noeqref{eq:14c}
\end{align} 
\label{eq:14}
\end{subequations}

\noindent Hence, when working with \eqref{eq:14}, learning $\cD$ is equivalent to learning $\widehat{\cD}$.   Once $\cW$ and $\cD$ in \eqref{eq:14} are learned from data using samples of $(\tensor{F}, \tensor{P}^\dagger)$, we obtain the trajectories of internal variables $\xi(t)$ as a by-product of the prediction of stress from strain. These internal variables are not uniquely identifiable given the dataset. In Section~\ref{sec:Int}, we present an analysis that asserts that the internal variables learned by \eqref{eq:14} from data are unique up to a linear transform.

\section{Methodology}
\label{sec:M}
In Section~\ref{ssec:GF}, we present a method to ensure material stability by incorporating growth in the learned constitutive law. Section~\ref{ssec:NNA} describes different architectures of neural networks that are used as building blocks in parameterizing $ \cW$ and $\cD$. Section~\ref{ssec:PNN} is the decomposition of the energy function into isochoric and volumetric components. This decomposition turns out to be beneficial for accurate prediction.  Details about the dataset and the NN training are presented, respectively, in Sections~\ref{ssec:data} and \ref{ssec:TNN}.

\subsection{Functions to Impose Material Stability}
\label{ssec:GF}
\begin{figure}[H]
    \centering
    \includegraphics[width=0.325\linewidth]{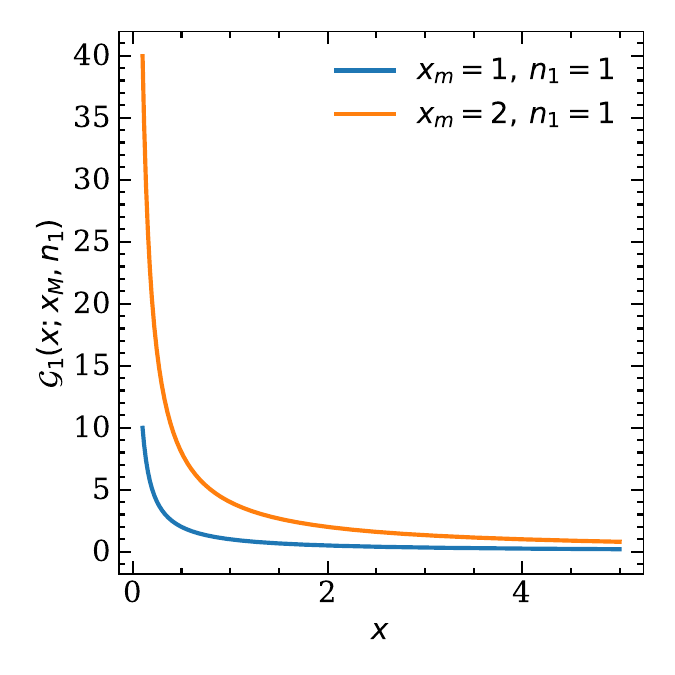}
    \caption{Visualization of $\mathcal{G}_1$ for different parametric values of $x_m$.}
    \label{fig:growth1}
\end{figure}

\begin{figure}[H]
     \centering
     \begin{subfigure}[b]{0.325\textwidth}
         \centering
         \includegraphics[width=\textwidth]{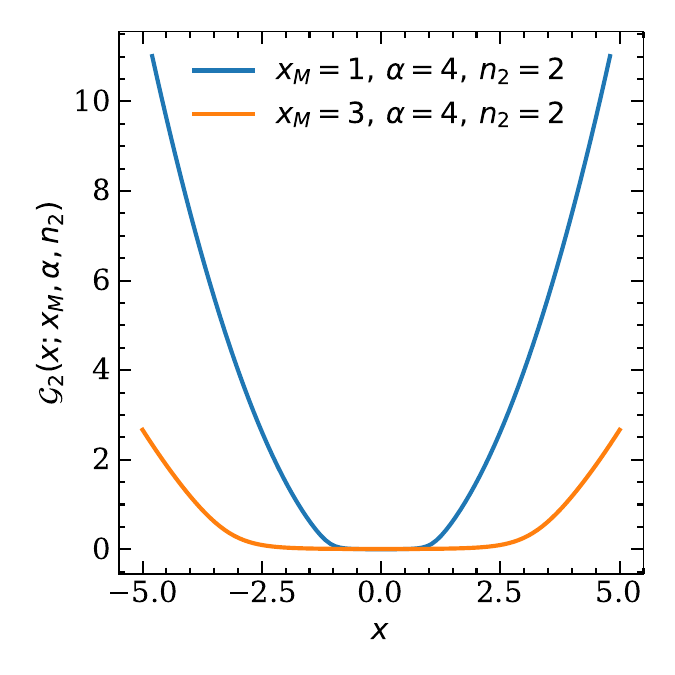}
     \end{subfigure}
     \hfill
     \begin{subfigure}[b]{0.325\textwidth}
         \centering
         \includegraphics[width=\textwidth]{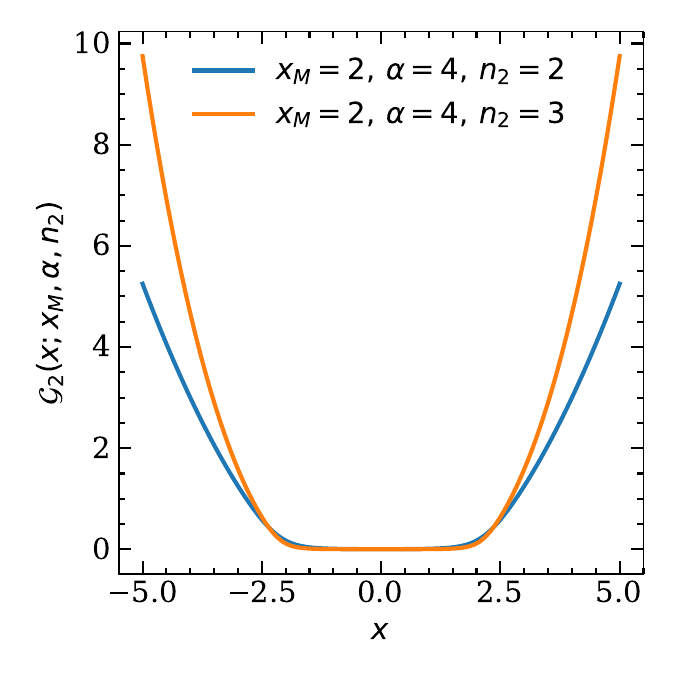}
     \end{subfigure}
     \hfill
     \begin{subfigure}[b]{0.325\textwidth}
         \centering
         \includegraphics[width=\textwidth]{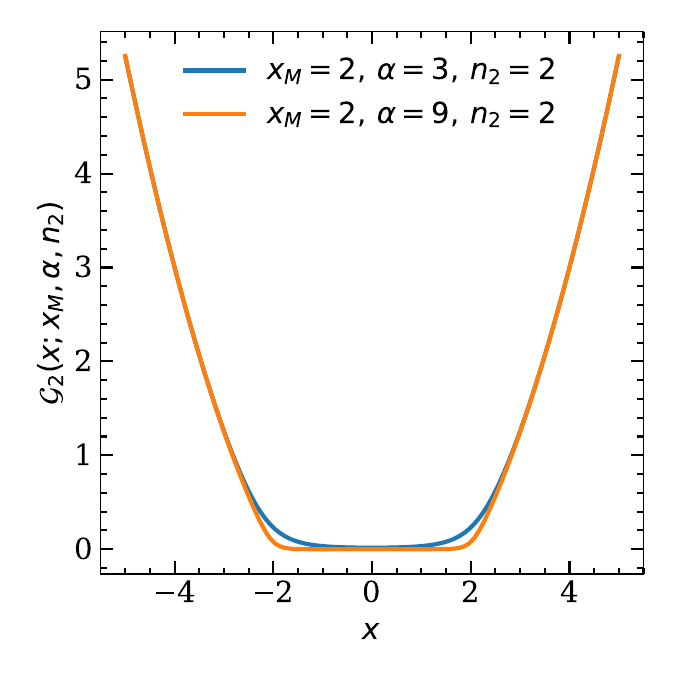}
     \end{subfigure}
        \caption{Effect of different parameters used in $\mathcal{G}_2$ on its gorwth.  $\mathcal{G}_2$ is used to enforce the growth of the potential outside the support of the data.
        }
        \label{fig:growth2}
\end{figure}

A NN can be chosen to approximate any continuous function arbitrarily well in the support of the dataset \citep{pinkus1999approximation}.  However, an optimized NN does not necessarily have the desired growth outside of the support of the dataset; this issue is addressed for dissipative fluid mechanics problems
in \citep{li2022learning}. Therefore, the growth of the energy potential, as discussed in Section~\ref{ssec:MCL}, cannot be learned by a NN from the data and must be incorporated separately. The two methods
proposed in \citep{li2022learning} to control the growth of a NN model are either (a) to augment the dataset with artificial data generated from an artificial model with desired properties (growth in our case), or (b) by adding functions to the NN model which enforce the desired properties without interfering with the learning process. We adopt the latter approach.
To this end, for $x \in \R$, the functions $\mathcal{G}_i, i=1,2$ are defined. These functions $\mathcal{G}_1:\R^+ \to \R$ and $\mathcal{G}_2: \R \to \R$ have the property that they are almost constant in the support of the dataset, but have polynomial growth outside the support.
We set 
\begin{align}
    \mathcal{G}_1(x; x_m, n_1) = \dfrac{1}{n_1}\dfrac{x_m^{n_1+1}}{x^{n_1}}; \quad x>0,\,\, x_m > 0
\end{align}
where 
\begin{align}
    x_m = \min_{x \in \text{Dataset}} x. \label{eq:x_m}
\end{align}
We note that, for $x_m$, the minimum must be taken over a positive valued quantity, denoted by the dummy variable $x$ in \eqref{eq:x_m}, such as the Jacobian of the deformation gradient, derived from the samples in the dataset. Real non-negative number $n_1 \in \R^+$ controls the singular behaviour at the origin.
We also define $\mathcal{G}_2:\R \to \R$ as
\begin{align}
    \mathcal{G}_2(x; x_M, \alpha, n_2) = \dfrac{x_M}{\alpha n_2} \log(1 + \exp (\alpha( \abs{x/x_M}^{n_2}-1)))
\end{align}
where 
\begin{align}
    x_M = \max_{x \in \text{Dataset}} \abs{x},
\end{align}
and $n_2 \in \R^+$ is the desired order of growth.  $x_M$ is a measure of the size of the data support, and the parameter $\alpha \in \R^+$,  which controls the transition to growth on the boundary of the support, needs to be appropriately chosen. 
We visualize the two functions $\mathcal{G}_i; i=1,2$. Figure~\ref{fig:growth1} shows the parametric dependence $\mathcal{G}_1(x; x_m, n_1)$ on $x_m$, for fixed $n_1$.  Figure~\ref{fig:growth2} shows parametric dependence of $\mathcal{G}_2(x; x_M, \alpha, n_2)$ on $x_M, \alpha$ and $n_2$. The parameter $x_M$ controls the size of the flat part of this function, $n_2$ is the order of singularity, and $\alpha$ controls how rapidly the function transitions from the flat part to its polynomial behavior.  Recall that the parameters $x_m$ and $x_M$ are chosen in such a way that the dataset lies in the relatively flatter part of $\mathcal{G}_1$ and $\mathcal{G}_2$. This ensures that $\mathcal{G}_1$ and $\mathcal{G}_2$ enforce the desired growth of potentials without interfering with the learning, that happens in the support of the dataset, of NNs $\mathcal{N}_H$ and $\mathcal{N}_V$. We show the use of these functions leads to physically desirable predictions that concern the staibility of a material under extreme values of applied strain $\tensor{F}$ in Section~\ref{sssec:VOG}.

Instead of adding the $\mathcal{G}_i$, $i=1,2$, to the NNs, it is computationally beneficial and more accurate to add their derivatives to the appropriate stresses derived from the NNs; this enables us to bypass the need for automatic differentiation. To this end, the derivatives of $\mathcal{G}_i,\, i=1,2$, denoted respectively by $\cH_1: \R \to \R$ and $\cH_2:\R \to \R$ are given as 
\begin{subequations}
    \begin{align}
        \mathcal{H}_1(x;x_m, n_1) &:= \dfrac{\partial \mathcal{G}_1(x;x_m, n_1)}{\partial x} =  -\bkt{\dfrac{x_m}{x}}^{n_1+1};\\
        \mathcal{H}_2(x;x_M, \alpha, n_2)&:=\dfrac{\partial \mathcal{G}_2(x; x_M, \alpha, n_2)}{\partial x} = \dfrac{\abs{x}^{n_2-2}x}{x_M^{n_2-1}(1 + \exp (-\alpha (\abs{x/x_M}_2^{n_2}-1)))}.
    \end{align}
\end{subequations}We will use the notational shorthands $\kappa$ and $\Xi$ to  denote $(x_m, n_1)$ and $(x_M, \alpha, n_2)$, respectively. The functions described here are used , along with NNs, to parameterize energy and dissipation potential $\cW$ and $\cD$. Before we discuss the exact parameterization in section~\ref{ssec:PNN}, we present the NN architectures at an abstract level in Section~\ref{ssec:NNA}, that are used in the parameterization of potentials.  
\subsection{Neural Network Architecture}
\label{ssec:NNA}
In this Subsection, we discuss the NN architecture used in this work: ($\mathcal{A}_1$) a NN that is convex in terms of its arguments.


\textbf{$\mathcal{A}$. Input Convex Neural Network.} We use a shallow (meaning single hidden layer) architecture of the input convex NN described in \citep{amos2017input}. The architecture can be written as 
\begin{subequations}
\begin{align}
    h_1 &= \mathrm{ReLUS}\bkt{W_1h_0 + b_1};\\
    h_2 &= W_2h_1 + b_2. 
\end{align}
\end{subequations}
The elements of weight and bias matrices, $W_1$ and $b_1$, are real-valued, while those of $W_2$ are positive real numbers. The activation function $\mathrm{ReLUS}(\cdot)$ is the square of the ReLU activation function, i.e.,  for $x \in \R$$,  \mathrm{ReLUS}(x) = x^2\mathbbm{1}_{\{x>0\}}$. The positivity of the element of $W_2$ is ensured by writing it as the element-wise square of another matrix,  i.e, $W_2 = \widetilde{W}_2 \odot \widetilde{W}_2$, where $\widetilde{W}_2$ takes values in a real space. It is to be noted that the positivity of $W_2$ is intended to enforce convexity of the architecture.  We also highlight that the choice of activation function, $\mathrm{ReLUS}$ compared to the popularly used $\mathrm{ReLU}$ is because the architecture, in this work, is used to parameterize potentials; samples of which are not available. Hence, these NNs are trained on samples of their derivatives, which are stresses. Therefore, this choice of activation function ensures that the gradients of the loss function, which will be in terms of the double derivatives of the activation function, do not vanish.  Moreover, noting that the $\mathrm{ReLUS}(\cdot)$ has quadratic growth, the activation function limits the architecture to have just one hidden layer; as the growth of the NN would be $2^{L-1}$ when $L$ hidden layers are used, consequently, making the training of such architectures next to impossible due to blow-ups during their optimization.


Architectures described here will be used to parameterize functions $\cW$, and $\cD$, described in Section~\ref{ssec:PNN}.

\subsection{Polyconvex  Energy Function}
\label{ssec:PNN}
 We parameterize the potential $\cW$,  in the following manner to enforce polyconvexity. We adopt a hydrostatic-deviatoric decomposition of $\cW$. To this end, we have
    \begin{align}
        \cW(\tensor{F}, \xi) = \cW_{H}\bkt{I_F} + \cW_V\bkt{\tensor{K}, \xi}
        \label{eq:decomposition}
    \end{align}
where $I_F = \det \tensor{F}$ and $\tensor{K} := \tensor{F}/I_F^{1/3}$; $\cW_H: \R \to \R$, $\cW_V: \R^{d\times d} \times \R^n \to \R$. 
We define $\widetilde{\tensor{P}} := \tensor{PF}$, and then, straightforward calculation leads to 
\begin{subequations}
\begin{align}
     \mathrm{hyd}\, \widetilde{\tensor{P}} &= I_F \dfrac{\partial \cW_H(I_F
)}{\partial I_F} \mathbf{I}_3, \label{eq:hydP} \\
\mathrm{dev}\, \widetilde{\tensor{P}} &= \dfrac{1}{I_F^{1/3}}\bkt{\dfrac{\partial \cW_V(\tensor{K}, \xi)}{\partial \tensor{K}} \tensor{F} -\bkt{\dfrac{\partial \cW_V(\tensor{K}, \xi)}{\partial \tensor{K}}\cdot  \tensor{F}} \mathbf{I}_3}.\label{eq:devP}
\end{align}
\end{subequations}
To incorporate limiting behaviour, we further decompose $\cW_H$ and $\cW_V$ as 
\begin{subequations}
\label{eq:poly_model}
    \begin{align}
        \cW_H(I_F) &\coloneqq \cN_H(I_F) + \mathcal{G}_1(I_F; \kappa_2) + \mathcal{G}_2(I_F, \Xi_3), \label{eq:wh_decomposition}\\
        \cW_V(\tensor{K}, \xi) &\coloneqq \cN_V(\tensor{F}, \mathrm{adj}\, \tensor{F}, I_F, \xi) +  \mathcal{G}_2(\vert n_\tensor{F}\vert; \Xi_4); \label{eq:wv_decomposition}
    \end{align}
\end{subequations}
where $n_\tensor{F} \coloneqq (\tensor{F},\, \text{adj}\, \tensor{F}, I_F)$ represents vectorized concatenation; $\cN_H:\R \to \R$ and $\cN_V:\R^{d\times d}\times \R^{d\times d} \times \R^+ \times \R^n \to \R$. $\cW_H$ is trained on the samples of

Polyconvexity of $\cW(\tensor{F}, \xi)$ in terms of $\tensor{F}$ is ascertained by using convex NN architecture, $\mathcal{A}$, for both $\cN_H$ and $\cN_V$.  $\cN_H$ is trained on samples of $(I_F, \mathrm{hyd}\, {\widetilde{\tensor{P}}^\dagger})$, while $\widetilde{\mathcal{N}}_V$ is trained on samples of $(\tensor{F}, \mathrm{dev}\, \widetilde{\tensor{P}}^\dagger)$. The dissipation potential is parameterized as follows 
\begin{align}
    \cD(\dot{\xi}) = \cN_D(\dot{\xi}) + \cG_2(\dot{\xi}; \Xi_5),
\end{align}
where $\cN_D$ is parameterized using architecture $\cA_2$ and $\cG_2$, with approprite parameters $\Xi_5$, is used to for coercivity of $\cD$. 


\subsection{Data}
\label{ssec:data} The method used to generate data for this study is ad-hoc but based on physical intuition. It is to be noted that internal variables $\xi$ are physically unobserved, and hence the dataset comprises only deformation gradient and stress trajectories. We denote the dataset as 
\begin{align}
\mathrm{D} = \cbkt{(\tensor{F}^{(i)}, \tensor{P}^{\dagger(i)})}_{i=1}^N
\end{align}
where, for a natural number $M$, $\tensor{F}^{(i)}: \cbkt{0, \dots, M} \to \Rd$ and $\tensor{P}^{\dagger(i)}: \cbkt{0, \dots, M} \to \Rd$ are discretized trajectories of deformation gradient and stress.  It is to be noted that the time step associated with the dataset is chosen to be $\Delta t = 1/M$. The data comes from solving a finer-scale crystal plasticity model of a representative volume element (unit cell). Each unit cell is a three-dimensional polycrystalline cube with 128 randomly oriented grains. Each grain has three basal, three prismatic, and six pyramidal slip systems, in addition to six tensile twin systems (treated as pseudoslips). The material parameters were chosen to represent magnesium. The homogenized stress is obtained by using Taylor's averaging, where the solution of the equilibrium equation is replaced by the assumption that the deformation gradient is uniform across the unit cell
\citep{kocks2000texture}. Details of the data generation can be found in \citep{LIU2022104668}. This dataset constitutes an interesting numerical homogenization problem for two primary reasons. Firstly, the existence of any true potentials $\cW^\dagger$ and $\cD^\dagger$ for the homogenized response is not established; so finding them computationally provides
insight. Secondly, the dimensionality, $n$, and samples of internal variables are unknown, and again computation provides insight. The dataset is decomposed into
mutually disjoint train and test sets.

\subsection{Training the Neural Networks} 
\label{ssec:TNN}
Here we present details pertaining to training the NNs. The function $\cW$ and $\cD$ describing the mapping from $\tensor{F}$ to $\tensor{P}$ in \eqref{eq:ds1}, are to be learned through the NNs $\mathcal{N}_H,  \mathcal{N}_V$ and $\cD$ described in Section~\ref{ssec:PNN}. The NN  $\mathcal{N}_H$, concerning only the hydrostatic stress that is independent of the strain-history and hence of the internal variable $\xi$, is trained on the input-output pairs $(I_3, \mathrm{hyd}\, \widetilde{\tensor{P}}^\dagger)$, the mapping between them is given in \eqref{eq:hydP}.  To this end, using ~\eqref{eq:hydP}, the loss  is defined as 
\begin{subequations}
     \begin{align}
     \mathcal{L}_1(\Theta_1) &= \dfrac{1}{NM}\sum_{n = 1}^{N}  \sum_{m=0}^M\bkt{\mathrm{hyd}\,\widetilde{\tensor{P}}_{m}^{\dagger(n)} - \mathrm{hyd}\,\widetilde{\tensor{P}}_{m}^{(n)}}^2
 \end{align}
\end{subequations}
where $n$ and $m$ are respectively indexing over trajectories and time steps. 
The optimum value, $\Theta_1^*$, of the parameter is found by minimizing $\cL_1$ i.e., 
\begin{align}
    \Theta^*_1 = \argmin_{\Theta_1} \mathcal{L}_1(\Theta_1). \label{eq:opt1}
\end{align} 
 Notably, the contribution of growth functions $\mathcal{H}_1$ and $\mathcal{H}_2$ in stress is included in the loss $\cL_1$. Now, we describe learning the NN $\cN_V$, which, in contrast with $\cN_H$, is coupled with the NN $\cD$ through the ODE  describing the dynamics of the internal variables. Hence, training the NNs $\mathcal{N}_V$ and $\cD$ requires, firstly, solving a dynamical system and then optimizing their parameters. Using equations \eqref{eq:14} and \eqref{eq:wv_decomposition}, the evolution in time of internal variable $\xi$ and the deviatoric stress $\mathrm{dev}\,\widetilde{\tensor{P}}$ is given as 
\begin{subequations}
\label{eq:sd_dynamics}
 \begin{align}
     \mathrm{dev}\, \widetilde{\tensor{P}} &=\dfrac{\partial \cN_V(\tensor{F}, \mathrm{adj}\, \tensor{F}, I_F, \xi)}{\partial \tensor{F}} + \dfrac{\partial \cG_2(\abs{n_F}; \Xi_4)}{\partial \tensor{F}}\label{eq:sd_dynamics_1}\noeqref{eq:sd_dynamics_1}\\ 
     \dot{\xi}(t) &= \dfrac{\partial \cD(d; \Theta_3)}{\partial d},\quad
     \text{where } d = -\dfrac{\partial \mathcal{N}_V(\tensor{F}, \mathrm{adj}\, \tensor{F}, I_F, \xi; \Theta_2)}{\partial \xi};\quad \xi(0) = 0. \label{eq:sd_dynamics_2}
 \end{align}
 \end{subequations}
The loss corresponding to $\mathrm{dev}\,\widetilde{\tensor{P}}$ is computed as 
 \begin{align}
     \mathcal{L}_2(\Theta_2, \Theta_3) = \dfrac{1}{MN}\sum_{n=1}^N\sum_{t=0}^M \norm{\mathrm{dev}\,\widetilde{\tensor{P}}^{\dagger(n)}_{m} - \mathrm{dev}\,\widetilde{\tensor{P}}^{(n)}_{m}}_{2}^2.
 \end{align}
We note that the the loss $\cL_2$ is a function of both $\Theta_2$ and $\Theta_3$, as $\mathrm{dev}\,\tensor{P}$ depends on both the NNs $\mathcal{N}_V$ and $\cD$. We also desire the convex NN $\cD(d, \theta_3)$ to have the minimum at zero. Therefore, we have a loss penalizing gradient at zero as follows 
\begin{align}
    \cL_3(\Theta_3) = \norm{\dfrac{\partial \cD(\mathbf{0}, \Theta_3)}{\partial d}}_2^2.
\end{align}
NNs $\cN_V$ and $\cD$ are optimized as 
\begin{align}
    \Theta_2^*, \Theta_3^* = \argmin_{\Theta_2, \Theta_3} (\cL_2(\Theta_2, \Theta_3) + \cL_3(\Theta_3)) \label{eq:opt2}
\end{align}
In both the optimization tasks, \eqref{eq:opt1} and \eqref{eq:opt2}, the parameters $\Theta_{1,2,3}$ are randomly initialized, the backpropagation algorithm is used to compute gradient of the loss with respect to the parameters, and Adam \citep{kingma2017} scheme is used to update the parameters. 
The training is implemented in PyTorch \citep{NEURIPS2019_PyTorch}.


Analogous procedure is followed for learning the polyconvex model by training ${\cN}_H, {\cN}_V$ and $\cD$ as per \eqref{eq:3}, \eqref{eq:poly_model}, based on appropriate input-output pairs. Once the NNs are trained, predictions of various quantities, such as stress $\tensor{P}$, can be done by evaluating the NNs with their parameters set to the optimum values, i.e., $\Theta_{1,2,3}^*$.
In the next Section, we present performance and evaluation of the this polyconvex model, empirical evidence of uniqueness of internal variables and supporting theory. 

\section{Results and Discussion}
\label{sec:rd}
The surrogate NN constitutive laws developed in this work are deployed to predict Taylor-averaged response of a three-dimensional magnesium polycrystal. Below, we list key results of this work and describe the order in which they are presented in this Section. The key results are as follows. 
\begin{enumerate}[label=(R\arabic*)]  
    \item Polyconvex neural network achieve relative error of 2\%  in prediction of Taylor-averaged response of a magnesium polycrystal, while being consistent with various desirable physical properties listed before in Section~\ref{sec:background}. The learned NN models also show the growth behavior achieved by using the functions $\mathcal{G}_{1,2}$. 
    \item The polyconvex model is learned on the dataset comprising only symmetric deformation gradients, and the model is extended to non-symmetric deformation gradients in a manner consistent with objectivity.
    \item The internal variables are empirically observed to be unique up-to linear transformation, and supporting theory is provided.  
\end{enumerate}
Section~\ref{ssec:DGFV} presents a qualitative analysis and visualization of the dataset. We then present the results corresponding to the polyconvex model in Section \ref{ssec:PNN}.
There, we show the predictive accuracy (R1) of the polyconvex model and visual comparisoin between true and predicted stress. In Section~\ref{ssec:OPM}, we present (R3) objective extension of the polyconvex neural network. It is followed in Section~\ref{sec:Int} by empirical evidence of that indentifiability of internal variables up to a linear transform, supporting theory, and the behaviour of the model under extreme stress.

\subsection{Data Visualization} \label{ssec:DGFV}
\begin{figure}[H]
     \centering
     \begin{subfigure}[b]{0.325\textwidth}
         \centering        \includegraphics[width=\textwidth]{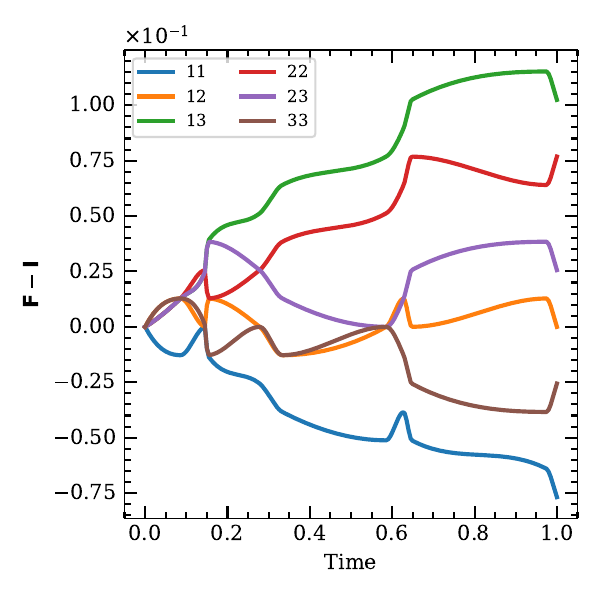}
     \end{subfigure}
     \begin{subfigure}[b]{0.325\textwidth}
         \centering
         \includegraphics[width=\textwidth]{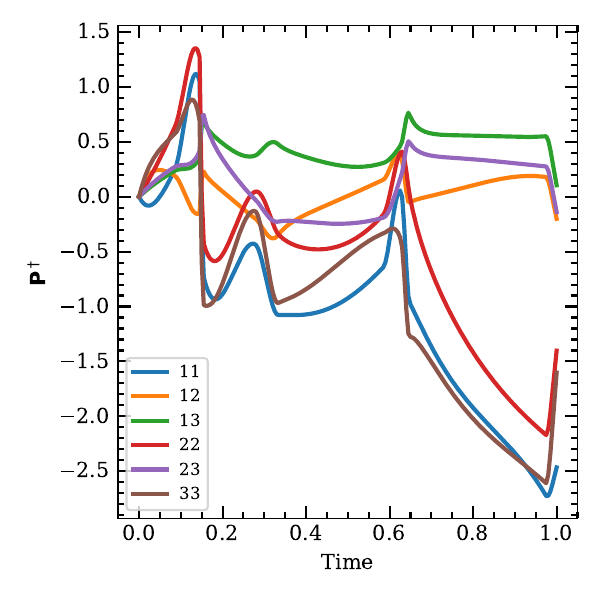}
     \end{subfigure}
        \caption{A sample from the dataset. There are six components of the deformation gradient tensor $\tensor{F}$ (left) as it is sampled to be symmetric and  (right) stress $\tensor{P}^\dagger$ (right).}
        \label{fig:sample_data}
\end{figure}

\begin{figure}[H]
     \centering
     \begin{subfigure}[b]{0.325\textwidth}
         \centering
         \includegraphics[width=\textwidth]{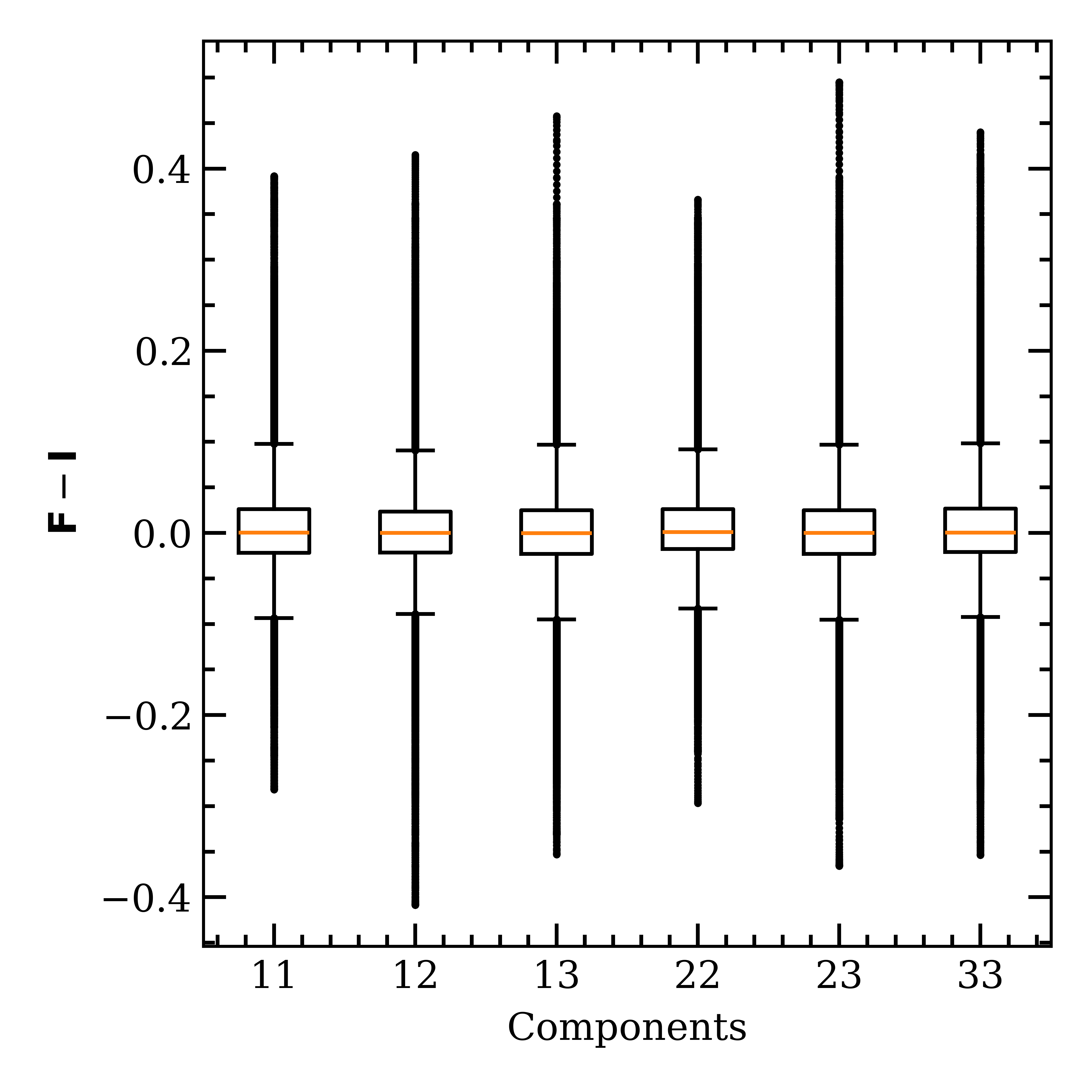}
     \end{subfigure}
     \hfill
     \begin{subfigure}[b]{0.325\textwidth}
         \centering
         \includegraphics[width=\textwidth]{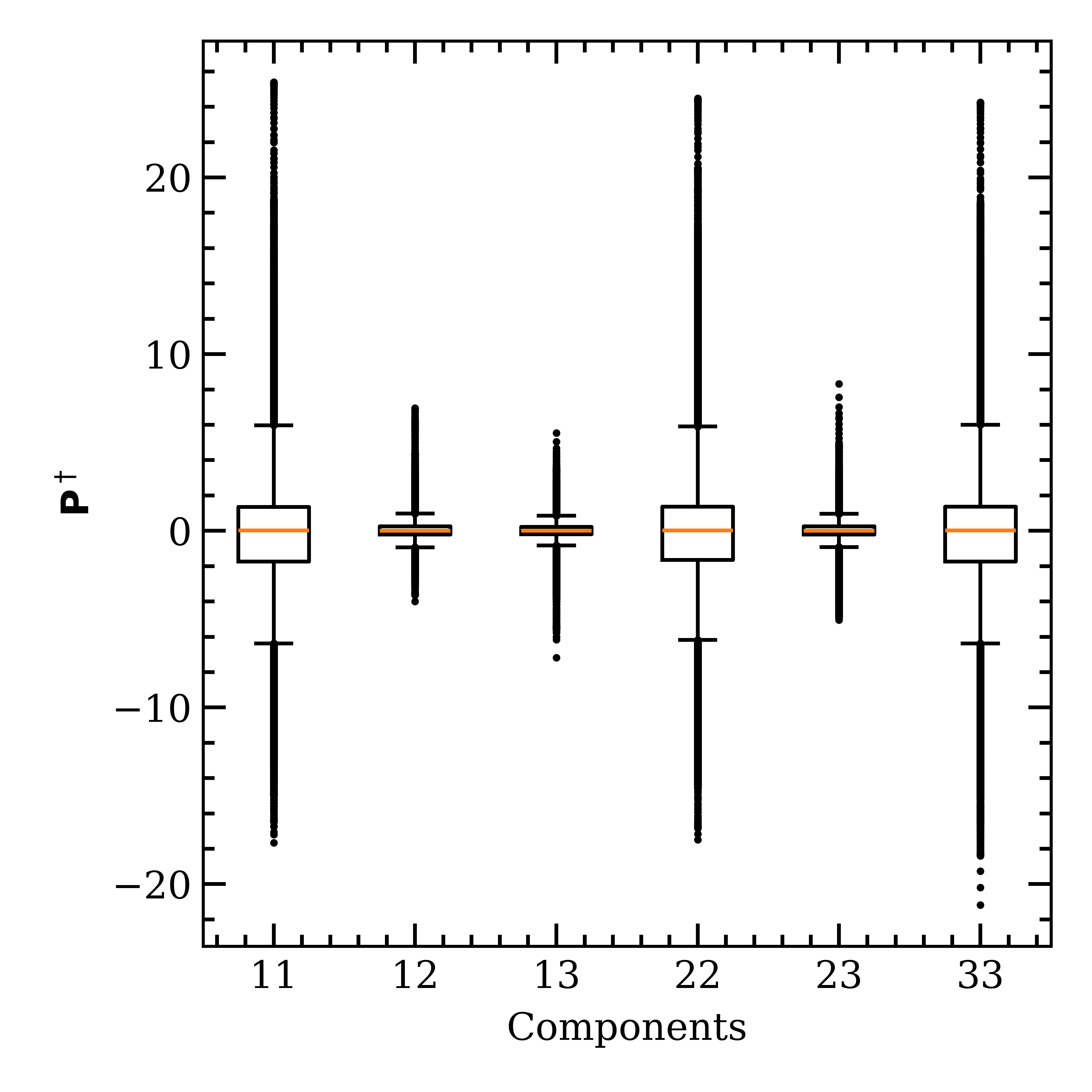}
     \end{subfigure}
     \hfill
     \begin{subfigure}[b]{0.325\textwidth}
         \centering
         \includegraphics[width=\textwidth]{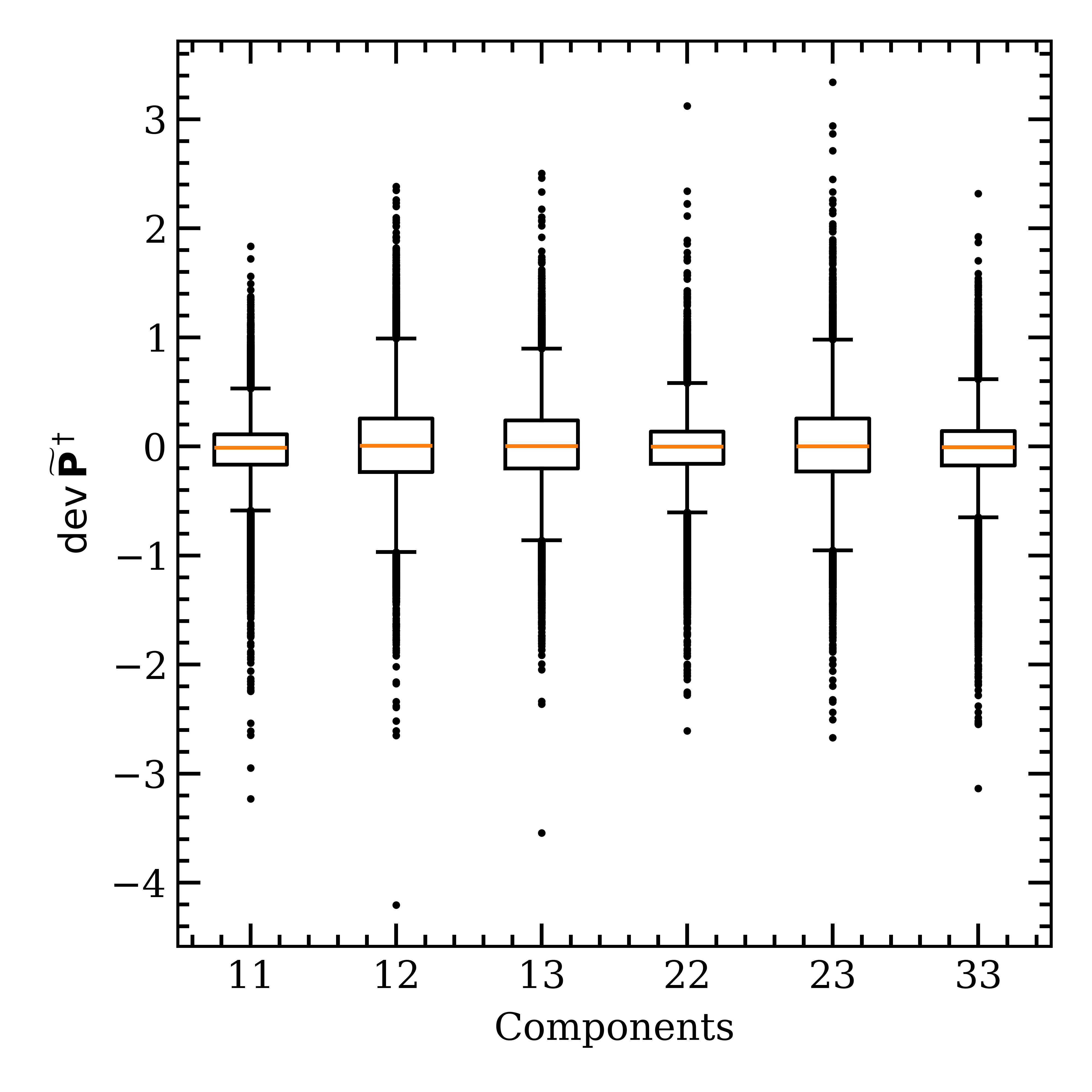}
     \end{subfigure}
        \caption{Distribution of different components of $\tensor{F}$,  $\tensor{P}^\dagger$ and $\tensor{P}^\dagger_d$.  There is more disparity in distributions of components of $\tensor{P}^\dagger$ as compared to those of $\mathrm{dev}\,\widetilde{\tensor{P}}^\dagger$.}
        \label{fig:data_dist}
\end{figure}

Figure~\ref{fig:sample_data} presents a sample from the dataset. Each sample constitutes components of the deformation gradient and stress tensor over a unit time interval. Though the deformation gradient tensor, $\tensor{F}$, is not necessarily symmetric, the dataset is generated by sampling $\tensor{F}$ to be symmetric.  The samples, each of $\tensor{F}$ and $\tensor{P}$ are shown in Figure~\ref{fig:sample_data}. 
\noindent We also visualize distributions of different quantities using box plots. It is observed from Figure~\ref{fig:data_dist} that the spread of different components of $\tensor{F}$, about their means, is comparable, as trajectory components are sampled in a similar manner. On the other hand, the diagonal components of stress $\tensor{P}$ have a much larger spread compared to the off-diagonal components.  If $\tensor{P}$ were to be learned directly by parameterizing the potential $\cW$ as a NN, without the decomposition~\eqref{eq:decomposition}, this disparity in the spread of stress components would make the pertinent training of the NN hard. Hence, $\cW$ is approximated by learning the  the NNs $\mathcal{N}_H$ and $\mathcal{N}_V$; where $\mathcal{N}_V$ is trained on samples of $\mathrm{dev}\,\widetilde{\tensor{P}}$. The right-most plot in Figure~\ref{fig:data_dist} shows the spread of different components of $\mathrm{dev}\,\widetilde{\tensor{P}}$, which are more comparable among themselves, as compared to that in $\tensor{P}$. This leads to enhanced predictive accuracy of this framework, which is presented later in Section~\ref{sssec:PSEA}, Figure~\ref{fig:error_analysis}.

\subsection{Polyconvex Model}
\label{ssec:OM}
In this section, we present results of the NN model described before in Section~\ref{ssec:PNN} that, by design, is polyconvex. 
Section~\ref{sssec:INIV} shows that six internal variables are sufficient to capture the history dependency of stress on strain. 
Section~\ref{sssec:PSEA} shows high predictive ability of the polyconvex model. We also show that the deviatoric component of the stress contributes significantly more to the overall error as compared to the hydrostatic component.  Section~\ref{sec:Int} shows that different sets of internal variables, obtained from two NN models differing in the initialization of their parameters at the start of training, are related through a linear transform. Section~\ref{sssec:VOG} shows that, as a result of using growth functions, the surrogate model exhibits physically desirable behaviour when subjected to extreme strain. 

\subsubsection{Identification of Number of Internal Variables} \label{sssec:INIV}
\begin{figure}[H]
    \centering    \includegraphics[width=\linewidth]{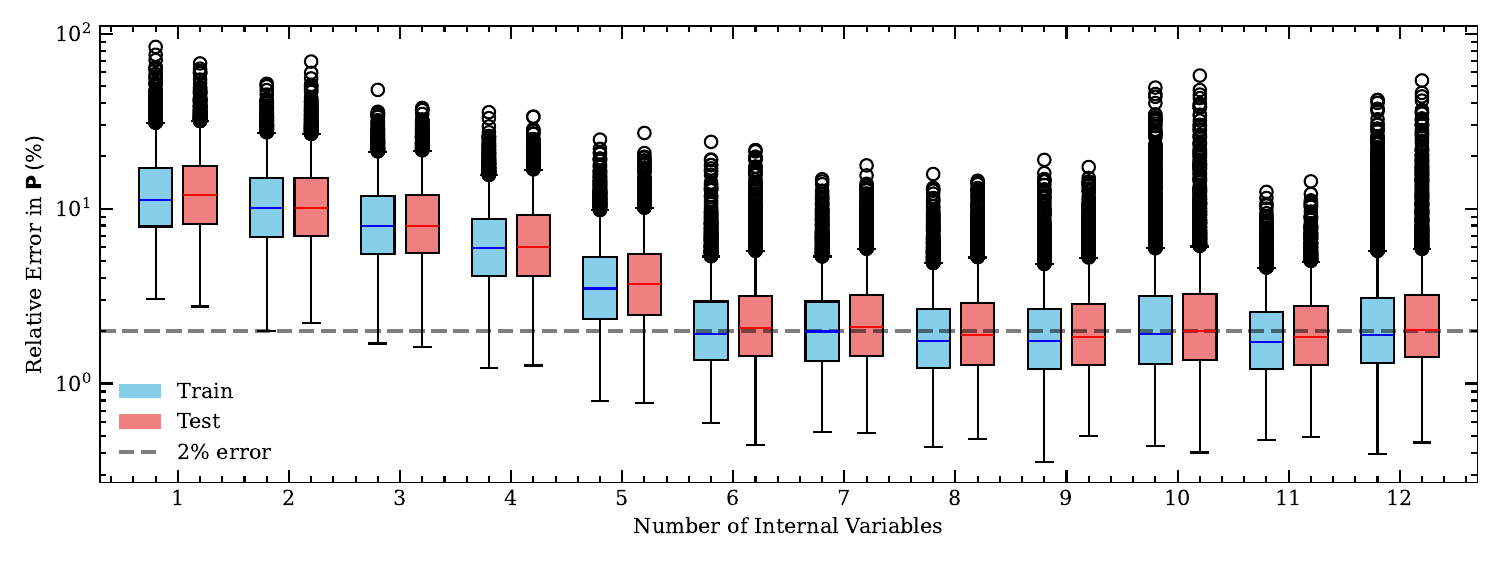}
    \caption{The use of six internal variables is the optimal choice for maximum prediction accuracy.}
    \label{fig:niv}
\end{figure}

Internal variables $\xi(t) \in \R^{n}$ are not physically observed quantities. They are introduced to approximate the non-Markovian dependence of $\tensor{P}$ on $\tensor{F}$ in a Markovian fashion. Therefore, the dimensionality, $n$, of internal variables is not known apriori; rather it is to be identified by learning NNs with different values of $n$ and choosen based on a balance between a desirable level of accuracy of the predicted $\tensor{P}$ with respect to $\tensor{P}^\dagger$ and dimension of the model. Figure~\ref{fig:niv} presents the relative error in predicted stress $\tensor{P}$ with respect to the dimensionality of the internal variables. All these NNs, corresponding to different values of $n$, were trained with their architecture and dataset size kept constant. $\mathcal{N}_V$ and $\cD$ have $15000$ nodes each in their only hidden layers. $\mathcal{N}_H$ has two hidden layers with $300$ nodes each. The numbers of input nodes in $\mathcal{N}_V$ and $\cD$ depend on $n$ and, hence, is equal to $6 + n$ and $n$ respectively. It may be observed from Figure~\ref{fig:niv} that $n=6$ is the optimal dimensionality of the internal variables for this numerical homogenization problem, a further increase in $n$ does not offer significant improvement in the prediction of stress, and the dimension $n=6$ does not impose an unwanted computational burden.

\subsubsection{Prediction of Stress and Error Analysis} \label{sssec:PSEA}
\begin{figure}[H]
    \centering 
    \begin{subfigure}{0.325\textwidth}
    \centering\includegraphics[width=\textwidth]{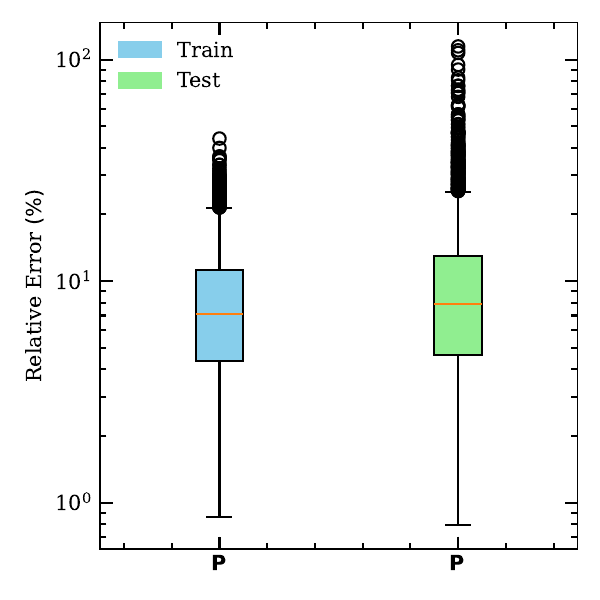}
\end{subfigure}
     \begin{subfigure}[b]{0.325\textwidth}
         \centering
         \includegraphics[width=\textwidth]{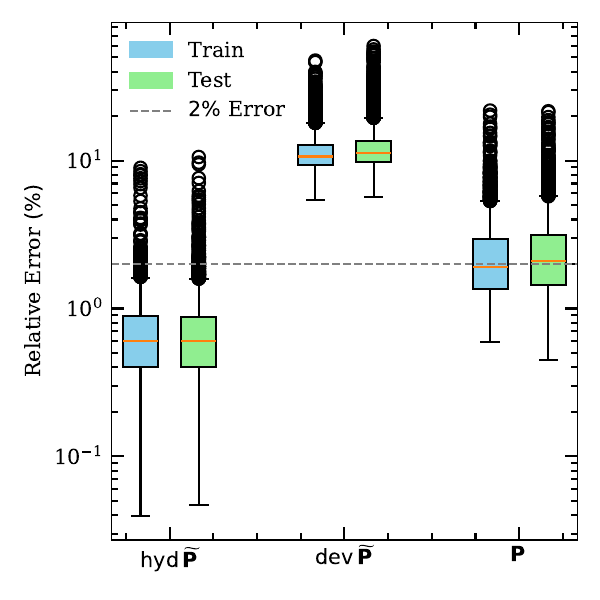}
     \end{subfigure}
    \caption{Relative error in stress. (Left) Training without decomposition of $\cW$ into its deviatoric-hydrostatic part. (Right) Training with decomposition of energy into deviatoric-hydrostatic parts as discussed in Section~\ref{ssec:PNN}. It is observed that the decomposition of energy results in significant improvement in the accuracy of the framework.}
    \label{fig:error_analysis}
\end{figure}

\begin{figure}[H]
    \centering    \includegraphics[width=\linewidth]{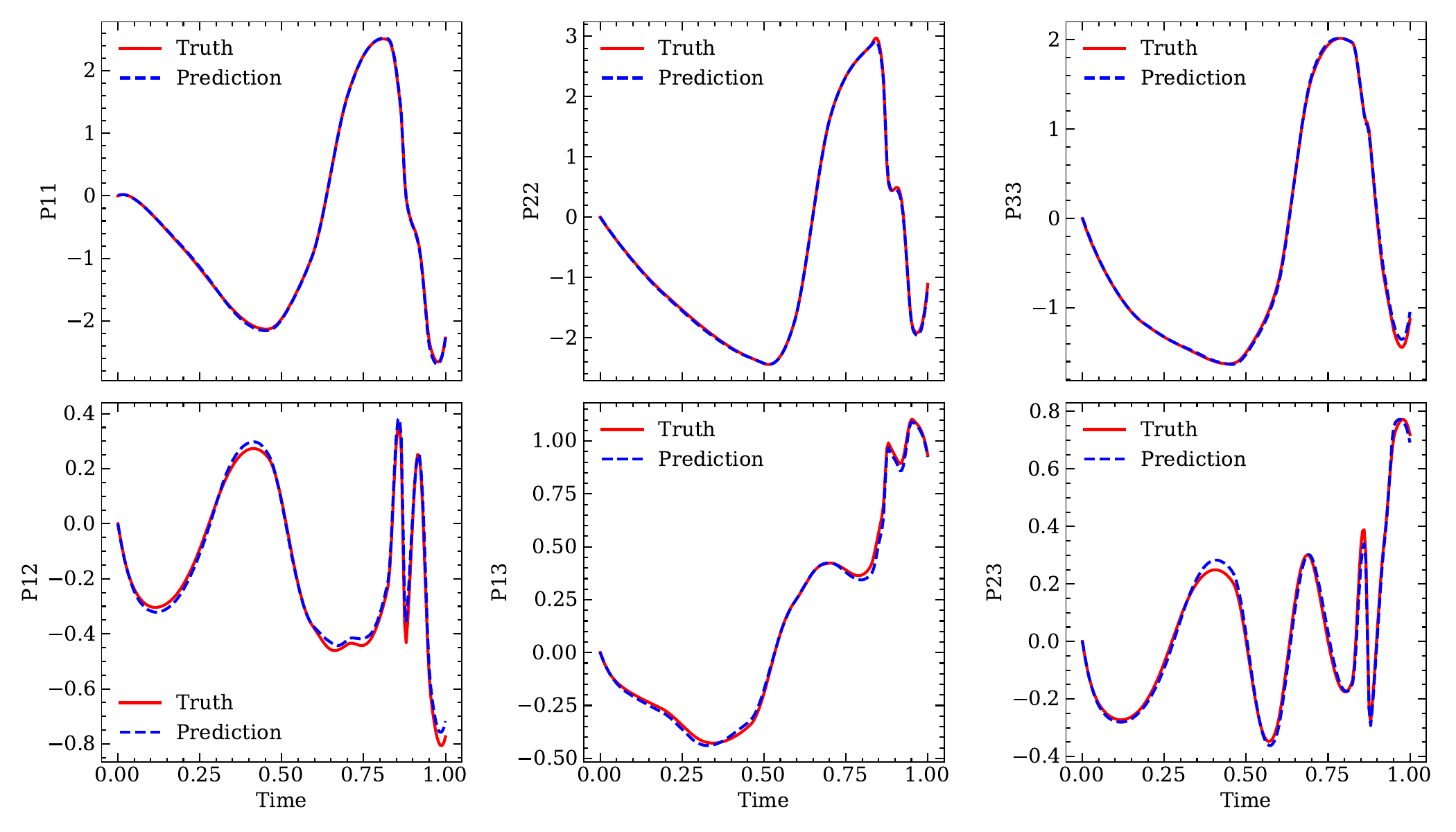}
    \caption{Visual comparison of different components of true and predicted stress. The sample corresponds to the median error of $2\%$. The plots in the top row show diagonal components of the stress tensor. The bottom row shows the off-diagonal components.}
    \label{fig:visual_comparison}
\end{figure}

After identifying the optimal dimensionality of internal variables and size of the NNs and dataset, we visualize the predictive accuracy of the objective framework and empirically analyze the error. The results correspond to the largest architecture that we trained. The number of nodes used in $\mathcal{N}_V$ and $\cD$ is $15000$ each and that in $\mathcal{N}_H$ is $300$. Three thousand samples are used for training and testing the NN model. The dimensionality of the internal variables used is six. Figure~\ref{fig:error_analysis} shows comparative performances of frameworks without and with deviatoric-hydrostatic decomposition of energy. We observe in Figure~\ref{fig:error_analysis} that the relative error in stress, in the case without decomposition, is 15\%, and that in the case with decomposition is 2\%. Moreover, when the energy is decomposed, we observe that the error in $\mathrm{hyd}\, \widetilde{\mathbf{P}}$ is much lower than that in $\mathrm{dev}\, \widetilde{\tensor{P}}$. The error in $\tensor{P}$ depends on the errors both in $\mathrm{hyd}\,\widetilde{\tensor{P}}$ and $\mathrm{dev}\, \widetilde{\tensor{P}}$. Though the error in $\mathrm{dev}\, \widetilde{\tensor{P}}$ is higher than that in $\mathrm{hyd}\, \widetilde{\tensor{P}}$, its contribution to the error is $\tensor{P}$ is not as much due to the lower magnitude of $\mathrm{dev}\,\widetilde{\tensor{P}}$ compared to that of $\mathrm{hyd}\, \widetilde{\tensor{P}}$. It is also observed that the errors on the training and test sets are close, reflecting the good generalization of the model. Figure~\ref{fig:visual_comparison} presents the visual comparison for a sample from the test dataset. The six plots compare the true stress components with their predicted values. 

\subsubsection{Visualization of Material Stability} \label{sssec:VOG}
\begin{figure}[H]
    \centering
    \includegraphics[width=0.8\linewidth]{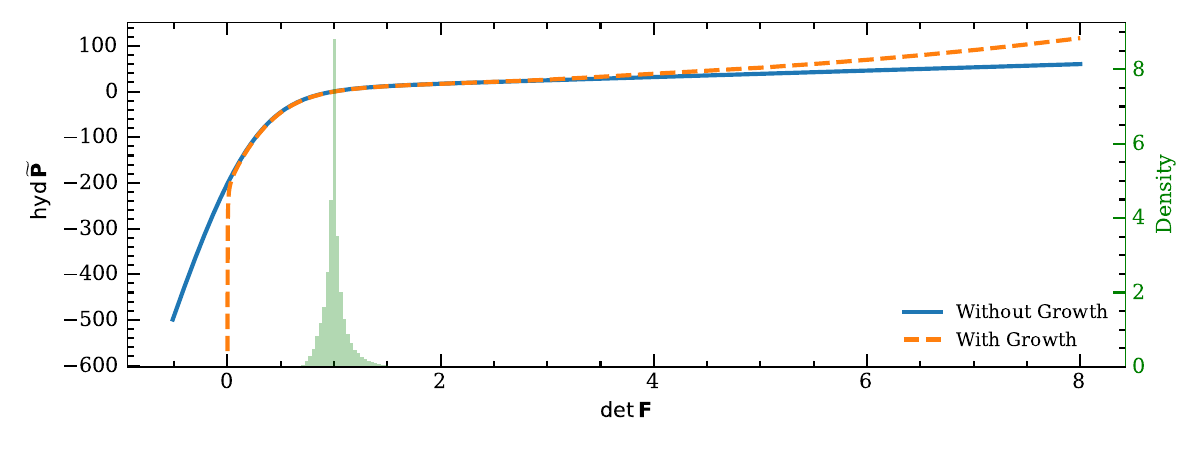}
    \caption{The hydrostatic stress with and without the growth imposed on the NN model for a fixed value of internal variables.}
    \label{fig:growth_comp1}
\end{figure}

In this Section, we present the effects of using growth functions $\mathcal{G}_i; i=1,2$, described in Section~\ref{ssec:GF},  in the prediction of stress for large strain values. Recall that $\mathcal{G}_i; i=1,2$ are used to enforce the physically reasonable property of any constitutive law that a piece of material cannot be compressed to a point with finite stress. Figure~\ref{fig:growth_comp1} shows the effects of using these functions with $\mathcal{N}_H$ (see~\eqref{eq:wh_decomposition}), the hydrostatic part of the model. It can be observed from Figure~\ref{fig:growth_comp1} that, if $\mathcal{G}_1$ is not used while training, i.e., growth is not imposed, the predicted stress is finite at $\det \tensor{F} = 0$. This is physically unreasonable, as it implies that a material can be compressed to a point with finite stress, hence would be called an unstable material. Therefore, this problem is rectified by using $\mathcal{G}_1$. From Figure~\ref{fig:growth_comp1}, it is observed that the hydrostatic stress $\mathrm{hyd}\,\widetilde{\tensor{P}}$ blows up to negative infinity as $\det \tensor{F}$ approaches zero from the right side. 

\begin{figure}[H]
    \centering
    \includegraphics[width=\textwidth]{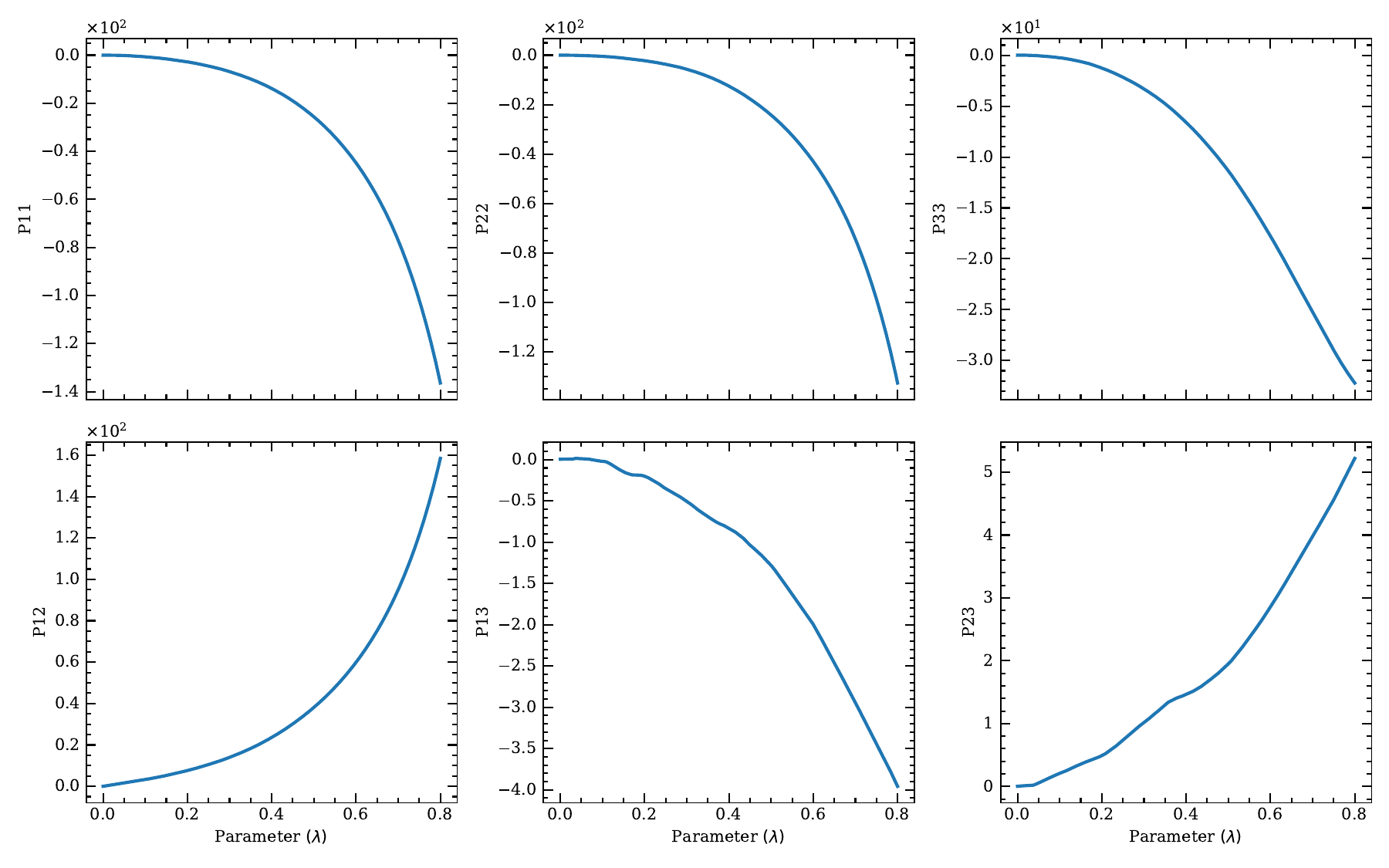}
    \caption{Increase in stress as shear strain is applied in the 1-2 plane, along the pure shear direction and beyond the support of the dataset (see~\eqref{eq:shear}).}
    \label{fig:pure_shear}
\end{figure}
\begin{figure}[H]
    \centering
    \includegraphics[width=\textwidth]{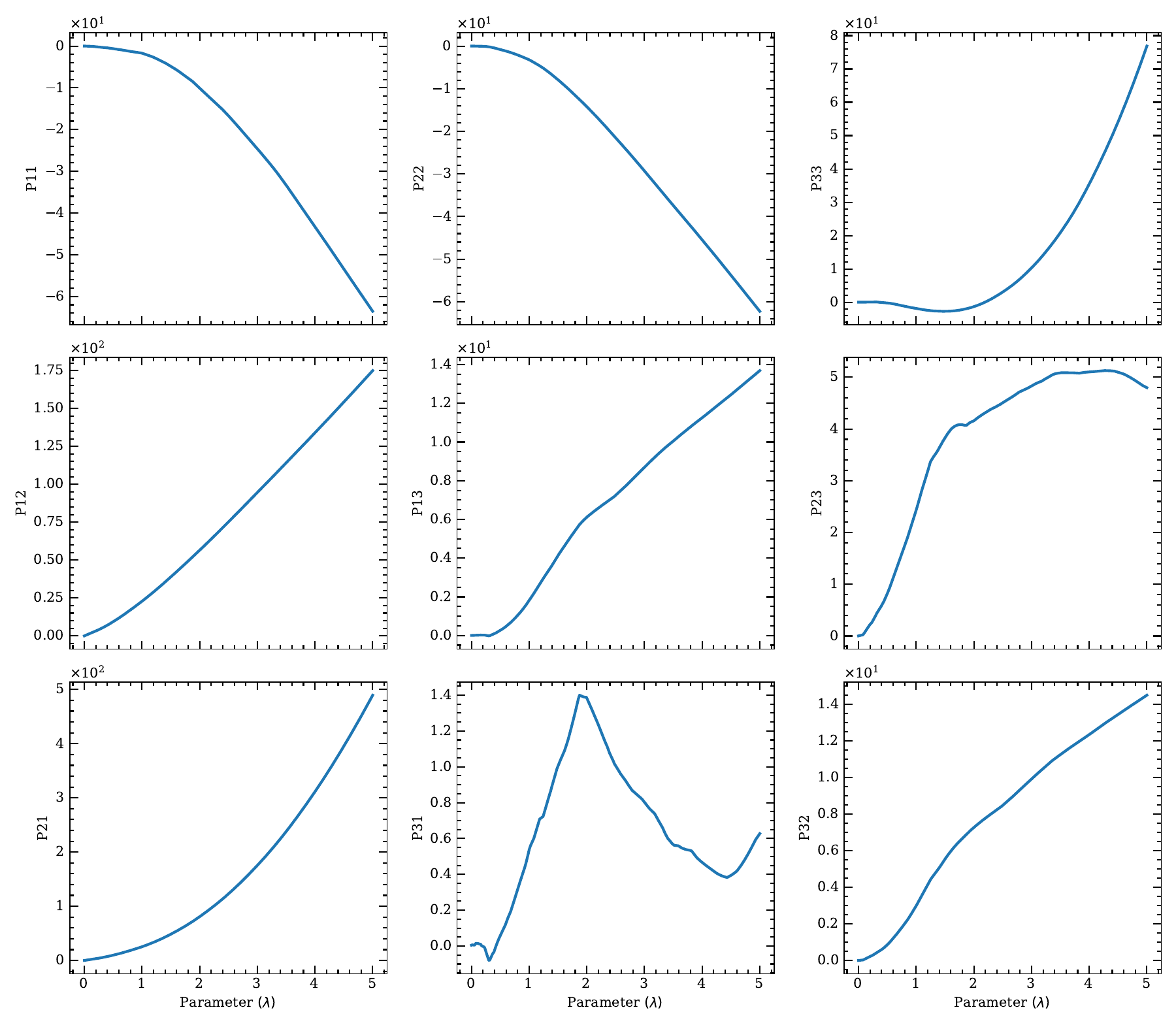}
    \caption{Increase in stress as shear strain is applied in the 1-2 plane, along the simple shear direction and beyond the support of the dataset (see~\eqref{eq:shear}).}
    \label{fig:simple_shear}
\end{figure}

Finally, we present the effect of using $\mathcal{G}_2$ with $\mathcal{N}_V$ (see~\eqref{eq:wv_decomposition}). To understand this, we examine the predicted stress for large shear strain. To this end, we fix the internal variable to be zero and predict stress along $\tensor{C}$ corresponding to pure shear and simple shear directions. Therefore, along pure shear and simple shear directions, we have
\begin{align}
    (\text{Pure shear}) &\quad\tensor{F}_\lambda = \begin{bmatrix}
        1 & \lambda & 0 \\
        \lambda & 1 & 0 \\
        0 & 0 &1
    \end{bmatrix}; \quad
    (\text{Simple shear}) \quad \tensor{F}_\lambda = \begin{bmatrix}
        1 & \lambda & 0 \\
        0 & 1 & 0 \\
        0 & 0 &1
    \end{bmatrix}.
    \label{eq:shear}
\end{align}
From Figure~\ref{fig:pure_shear}, we observe the desired growth in stress components for pure shear in the $1$-$2$ plane. Figure~\ref{fig:simple_shear} shows the growth of stress in a simple shear direction. From both Figure~\ref{fig:pure_shear} and Figure~\ref{fig:simple_shear}, we observe that normal components of stress along all three directions and shear components of stress in the $1-2$ plane grows rapidly with growing amount of shear, $\lambda$. The shear components $S13$ and $S23$, that do not lie in the $1-2$ plane, are almost small compared to the components in $1-2$ plane.  
\subsubsection{Objectivity of the Polyconvex Model}
\label{ssec:OPM}
We recall that the polyconvex model is trained on data comprising only symmetric deformation gradients. Hence, it is not objective by design of its architecture. The current literature shows that there are limited constitutive models of general forms that are both polyconvex and objective. While invariant-based neural-network models that satisfy both polyconvexity and objectivity have not performed well in terms of predictive accuracy, objectivity has been incorporated as a soft constraint by adding an additional term to the loss function \cite{KLEIN2022104703}. The term appended to neural network loss function for incorporating objectivity is as follows 
\begin{align}
   \norm{ \tensor{\tensor{P}(\tensor{QF}) - \tensor{Q}\tensor{P}(\tensor{F})}}^2, \quad \forall \tensor{Q}, \tensor{F}. \label{eq:objective_loss}
\end{align}
where $\tensor{Q}$ is an orthogonal matrix. In this work, we propose another method that is equivalent to appending \eqref{eq:objective_loss} to the loss term.

We recall that, in this work, the polyconvex neural network model is trained on dataset comprising only of symmetric deformation gradient $\tensor{F}$. Hence, the first Piola-Kirchoff stress $\tensor{P}_D$, where the subscript $D$ is used to represent map that is learned from data,  is accuracte over space of symmetric tensor i.e., $\tensor{P}_D: \R^{d \times d}_{sym} \to \R^{d\times d}_{sym}$. However, we extend this learned model to define stress for non-symmetric deformation gradient as follows
\begin{align}
    \tensor{P}(\tensor{F}) \coloneqq \begin{cases}
\tensor{P}_D(\tensor{F}),\quad &\text{if } \tensor{F} \in \R^{d\times d}_{sym}; \\
\widetilde{\tensor{Q}}\tensor{P}_D(\widetilde{\tensor{F}}), \quad &\text{if }\tensor{F} \neq \R^{d \times d}_{sym},\, \tensor{F} = \widetilde{\tensor{Q}}\widetilde{\tensor{F}}, \tensor{Q} \in \textrm{SO}(d),\, \widetilde{\tensor{F}} \in \R_{sym}^{d\times d}. 
    \end{cases}\label{eq:objectivity_definition}
\end{align}
$\tensor{F} = \widetilde{\tensor{Q}}\widetilde{\tensor{F}}$ is the polar decomposition of $\tensor{F}$. 
Hence, with \eqref{eq:objectivity_definition} extension of Piola-Kirchoff stress to non-symmetric deformation gradients, \eqref{eq:objective_loss} evaluates to zero as follows,
\begin{equation}
    \tensor{P}(\tensor{QF}) - \tensor{Q}\tensor{P}(\tensor{F}) = \tensor{P}(\tensor{Q}\widetilde{\tensor{Q}}\widetilde{\tensor{F}})-\tensor{Q}\tensor{P}(\widetilde{\tensor{Q}}\widetilde{\tensor{F}}) = \tensor{Q}\widetilde{\tensor{Q}}\tensor{P}_D(\widetilde{\tensor{F}}) - \tensor{Q}\widetilde{\tensor{Q}}\tensor{P}_D(\widetilde{\tensor{F}}) = \tensor{0}.
\end{equation}
\subsection{Identifiability of Internal Variables}
\label{sec:Int}
An important aspect of data-driven constitutive identification is the internal variables are not postulated a priori, but learned from stress-strain data.  Therefore, it happens that the same stress-strain data, but different instances of learning can lead to different internal variables.
In this section, we present the main theoretical results that internal variables are indentifiable upto linear transform. Throughout this subsection, we work with the constitutive law in \eqref{eq:ds1}, which has the internal variable evolution in implicit form.
Theorem~\ref{th:existence_of_solutions} states that given a form of constitutive model, which is defined by a fixed pair $(\cW, \widehat{\cD})$, as in \eqref{eq:10b}, and the strain-stress response generated by that model; a linear transformation of the internal variables does not change the strain-stress response. Theorem \eqref{th:main_th} states the converse result, i.e., given two pairs of functions $(\cW, \widehat{\cD})$ and $(\widetilde{\cW}, \widetilde{\cD})$ that lead to the same stress-strain response, and that differ only in their sets of internal variables, then the sets of internal variables must be related by a linear transform. This uniqueness of internal variables up to linear transform is empirically verified later in this section.
\begin{definition}
    Let $m, n$ be positive integers, $\R^m$ be the m-dimensional space of real numbers, and $\mathrm{GL}(n, \R)$ be the general linear group of order $n$. 
    For $T>0$, let $\cC^k=\cC^k([0, T]; \R^m)$ denote the class of $k$-times continuously differentiable functions, and $L^\infty([0, T]; \R^m)$ the essentially bounded functions, from $[0,T]$ into $\R^m.$
\end{definition}

\begin{assumption}
\label{assump:w_d}
    Let $\cW \in \cC^1(\R^m\times \R^n, \R^+)$. Let $\widehat{\cD} \in \cC^1(\R^n,  \R^+)$ be a strictly convex function with minimum value of zero attained at the point zero.
\end{assumption}
\begin{remark}
    In this section, $\cW$ is defined to take the vectorized deformation gradient as input. Hence, $m = d(d+1)/2$. We highlight that the $\cW$ is viewed as a function from $\R^m \times \R^n$, where  $\R^n$ is the space internal variable lives. We ignore the fact that the deformation gradient must have positive determinant. Morever, we note that the rate independent theories of plasticity are not strictly convex.
\end{remark}

\begin{definition} Let $\cS: \R^m \times \R^n \to \R^m$ and $\cF: \R^m \times \R^n \times \R^n \to \R^n$ be such that for $f \in \R^m$ and $p, q\in \R^n$
\begin{subequations}
\label{eq:seq}
\begin{align}
    \cS(f,p) &\coloneqq \dfrac{\partial \cW(f,p)}{\partial c},\label{eq:seqa}\\
    \cF(f, p, q) &\coloneqq \dfrac{\partial \widehat{\cD}(q)}{\partial q}+\dfrac{\partial \cW(f, p)}{\partial p}. \label{eq:seqb}
    \end{align}
\end{subequations}
\end{definition}
\begin{definition}  
    Let $C \in L^\infty([0,T]; \R^m)$ and $\xi \in \cC^1([0,T]; \R^n)$. For time $t \in [0, T]$, using \eqref{eq:seqa}, and \eqref{eq:seqb}, we define a first-order ordinary differential equation and an observable $S$ as follows
    \begin{subequations}
    \label{eq:sode}
        \begin{align}
    S(t)  &= \cS(F(t), \xi(t)) ,\label{eq:sodea}\\
    \mathcal{F}(F(t), \xi(t), \dot{\xi}(t)) &= 0,\quad \xi(0) = 0.\label{eq:sodeb}
    \end{align}
    \end{subequations}
\end{definition}


\begin{theorem}
\label{th:existence_of_solutions}
Given a pair of functions ($\cW$, $\widehat{\cD}$) satisfying Assumption~\ref{assump:w_d}. For any $F \in L^\infty$, consider  $S(t)$ and  $\xi(t)$ given by $\eqref{eq:sode}$. Then,  for every $A \in \mathrm{GL}(n, \R)$, there exist functions
\begin{subequations}
    \begin{alignat}{3}
        \widetilde{\cW}&:\R^m \times \R^n \to \R^+,\quad&(f, \widetilde{p}) &\mapsto \widetilde{\cW}(f, \widetilde{p});\\
        \widetilde{\cD} &: \R^n \to \R^+,\quad  &\widetilde{q} &\mapsto \widetilde{\cD}(\widetilde{q})
    \end{alignat}
\end{subequations}
    with $\widetilde{\cS}: \R^m \times \R^n \to \R^m$ and $\widetilde{\cF}: \R^m \times \R^n \times \R^n \to \R^n$, defined by,  
    \begin{subequations}
    \label{eq:stildeeq}
        \begin{align}
    \widetilde{{\cS}}(f, \widetilde{p}) &\coloneqq \dfrac{\partial \widetilde{\cW}(f, \widetilde{p})}{\partial c} \label{eq:stildeeqa}\noeqref{eq:stildeeqa}
\\
            \widetilde{\cF}(f, \widetilde{p}, \widetilde{q}) &\coloneqq \dfrac{\partial \widetilde{\cD}(\widetilde{q})}{\partial \widetilde{q}}+\dfrac{\partial \widetilde{\cW}(f, \widetilde{p})}{\partial \widetilde{p}} \label{eq:ftildeeqb}\noeqref{eq:ftildeeqb}
        \end{align}
    \end{subequations}
such that, for $\eta(t) := A\xi(t)$ for $t \in [0, T]$,
\begin{subequations}
        \begin{align}
    S(t) &= \widetilde{\cS}(F(t), \eta(t)); \\
    \widetilde{\cF}(F(t), \eta(t), \dot{\eta}(t)) &= 0,\quad \eta(0) = 0.
    \end{align} 
\end{subequations}
Moreover, the convexity of $\widehat{\cD}$ is preserved in $\widetilde{\cD}$.
\begin{proof}
    Proof given in the Appendix. 
\end{proof}
\end{theorem}
\begin{remark}
    Theorem~\ref{th:existence_of_solutions} holds for any positive integers $m$ and $n$, i.e., any dimensionality of $F(t)$ and $\xi(t)$.  
\end{remark}
Now, we state Theorem~\ref{th:main_th}, which is the converse of Theorem~\ref{th:existence_of_solutions}. 
\begin{definition}
For a given pair of functions $(\cW, \widehat{\cD})$, $t \in [0, T]$, $C \in L^\infty$, and the corresponding differential equation \eqref{eq:sodeb}, we define   
\begin{subequations}
\begin{align}
U &\coloneqq \{F(t): F \in L^{\infty}, t\in [0, T]\}\subset \R^m,\\
\Upsilon &\coloneqq \{\xi : \cF(F(t), \xi(t), \dot{\xi}(t))=0 ,\,\ \forall F \in L^\infty\} \subset \cC^1([0,T]; \R^n),\\
P &\coloneqq \{ \xi(t): \xi \in \Upsilon, t \in [0, T]\} \subset \R^n,\\
Q &\coloneqq \{\dot{\xi}(t): \xi \in \Upsilon, t \in [0, T]\} \subset \R^n.
\end{align}
\end{subequations}
\end{definition}

We also observe that sets $P$ and $Q$ are directly related to set $\Upsilon$. We desire $P$ and $Q$ to be open subsets of $\R^n$. To this end, we state Assumption~\ref{assup:controllability} needed for Theorem~\ref{th:main_th}. We later provide examples with particular forms of functions of $(\cW, \widehat{\cD})$ that are consistent with Assumption~\ref{assup:controllability}. 

\begin{assumption}\label{assup:controllability} For a given pair of functions $(\cW, \widehat{\cD})$, consider the ODE defined in \eqref{eq:sodeb}. Let there be open sets  $P, Q  \subset \R^n$ such that for any $(p, q) \in P \times Q$, there exists $t_* \in [0, T]$, $F_* \in L^\infty([0,T];\R^m)$ and $\xi_* \in \cC^1([0,1];\R^n)$ such that
\begin{subequations}
    \begin{align}
    \cF(F_*(t), \xi_*(t), \dot{\xi}_*(t)) &= 0,\quad \xi_*(0) = 0, \quad \forall t \in [0, T];\\
    \xi_*(t_*) = p; &\quad \dot{\xi}_*(t_*) = q.
\end{align}
\end{subequations}
\end{assumption}

\begin{definition} For $P \subset \R^n$,
   $\cT\in \cC^2(\R^n; \R^n)$ is a twice-differentiable diffeomorphism, which maps $P$ into $\R^n$, if and only if $\cT$ is bijective, and $\cT^{-1}$ is twice differentiable.
\end{definition}
\begin{theorem}
\label{th:main_th}
    For given two pairs of functions $(\cW, \widehat{\cD})$ and $(\widetilde{\cW}, \widetilde{\cD})$; consider ($\cS$,  $\cF$), and $(\widetilde{\cS}$, $\widetilde{\cF})$ defined as per \eqref{eq:seq}, and  \eqref{eq:stildeeq} respectively. Let for all $C \in L^\infty$ and $t \in [0, T]$,  
    \begin{subequations}
        \begin{align}
        \cS(F(t), \xi(t)) &= \widetilde{\cS}(F(t), \eta(t)),\\
        \cF(F(t), \xi(t), \dot{\xi}(t)) &= 0, \quad \xi(0) = 0, \\
        \widetilde{\cF}(F(t), \eta(t), \dot{\eta}(t)) &= 0, \quad \eta(0) = 0.
    \end{align}
    \end{subequations}
Moreover, let $\cF$ satisfy Assumption~\ref{assup:controllability} and $P \subset \R^n$ be the set as defined in Assumption~\ref{assup:controllability}. Let $\cT : P \to \R^n$ be a smooth diffeomorphism such that $\eta(t) = \cT(\xi(t))$. Then, $\cT$ must be  linear i.e. $\cT(\xi(t)) = A\xi(t)$ for all $\xi(t) \in P$, where $A \in \mathrm{GL}(n, \R)$. 
\begin{proof}
    The proof of Theorem~\ref{th:main_th} is given in the Appendix.
\end{proof}
\end{theorem}

 We now present a few examples to highlight the importance of the Assumption~\ref{assup:controllability} in Theorem~\ref{th:main_th}.
We present Example~\ref{ex:3} that is consistent with Assumption~\ref{assup:controllability}.
Example~\ref{ex:3} is a case where the dynamics of internal variables are governed by a non-linear ODE.  It is to be noted that the source of non-linearity is the choice of non-quadratic $\cD$; while $\cW$ is quadratic. 
\begin{example}
\label{ex:3}
Let $m = n$ and define $\cW$ and $\widehat{\cD}$  as follows 
    \begin{subequations}
    \begin{align}
        \cW(f, p) &= \dfrac{1}{2}\begin{bmatrix}
            f\\p
        \end{bmatrix}^\top \begin{bmatrix}
            A_1 & A_2\\
            A_2^{\top} & A_3
        \end{bmatrix} \begin{bmatrix}
            f \\ p
        \end{bmatrix},\\
        \widehat{\cD}(q) &= \dfrac{1}{2(r+1)}\abs{q}^{2(r+1)};\quad \text{ where } r> -1/2,  
    \end{align}
    \end{subequations}
where $\abs{\cdot}$ is the 2-norm.  
Hence, 
\begin{equation}
    \cF(F(t), \xi(t), \dot{\xi}(t)) = \abs{\dot{\xi}(t)}^{2r} \dot{\xi}(t) + A_3\xi(t) + A_2 F(t). 
\end{equation}
Hence, for any given $t_*$ and $p, q \in \R^n$, we can construct $\xi_*$ and $C_*$ as 
\begin{align}
    F_*(t) = -A_2^{-1}\bkt{\abs{\dot{\xi}_*(t)}^{2r} \dot{\xi}_*(t) + A_3\xi_*(t)}
\end{align}
such that $\cF(F_*(t), \xi_*(t), \dot{\xi}_*(t)) = 0$, $\xi_*(t_*) = p$ and $\dot{\xi}_*(t_*) = q$. Hence, $P=Q=\R^n$, and this example is consistent with Assumption~\ref{assup:controllability}.
\end{example}
Finally, we present a case to highlight the importance of dimensionality. In the following example, the dimensionality of the deformation gradient is less than that of the strain. This leads to the example being inconsistent with Assumption~\ref{assup:controllability}. Hence, the technique employed to prove Theorem~\ref{th:main_th} does not apply to Example~\ref{ex:4}. 

\begin{example}
\label{ex:4}Let $m < n$. Consider $\cW$ and $\widehat{\cD}$ as follows.
\begin{subequations}
    \begin{align}
        \cW(f, p) &= \dfrac{1}{2}\begin{bmatrix}
            f\\p
        \end{bmatrix}^\top \begin{bmatrix}
            A_1 & A_2\\
            A_2^{\top} & A_3
        \end{bmatrix} \begin{bmatrix}
            f \\ p
        \end{bmatrix},\\
        \widehat{\cD}(q) &= \dfrac{1}{2}q^{\top}Dq, 
    \end{align}
    \end{subequations}
where $D \in \R^{n\times n}$ is symmetric, $A_3 \in \R^{n \times n}$ is symmetric and $A_2 \in \R^{n \times m}$.
Hence, we have 
\begin{align}
    D\dot{\xi}(t) + A_3\xi(t) + A_2F(t) = 0,
\end{align}
     For a fixed $t = t_*$, fix $\xi(t_*) = p_*$. Hence, 
    \begin{align}
        \dot{\xi}(t_*) = - D^{-1}(A_3 p_* + A_2F(t_*)).
    \end{align}
    Define the set 
    \begin{align}
        Q \coloneqq \{\dot{\xi}(t_*) : F
        (t_*) = u, \, \forall u \in \R^m \}.
    \end{align}
    Since $A_2$ has rank of $m$, the set $Q \subset \R^n$ is m-dimensional. Hence, $\dot{\xi}(t_*)$ cannot take any arbitrary value in $\R^n$.
\end{example}
\begin{remark}
    The results about the identifiability of internal variables, established in this section, are agnostic of any structure, such as polyconvexity which is discussed in Section~\ref{ssec:PNN}, on $\cW$, except for the ones that might arise to satisfy Assumption~\ref{assup:controllability}. 
\end{remark}

\begin{figure}[H]
    \centering
    \includegraphics[width=\linewidth]{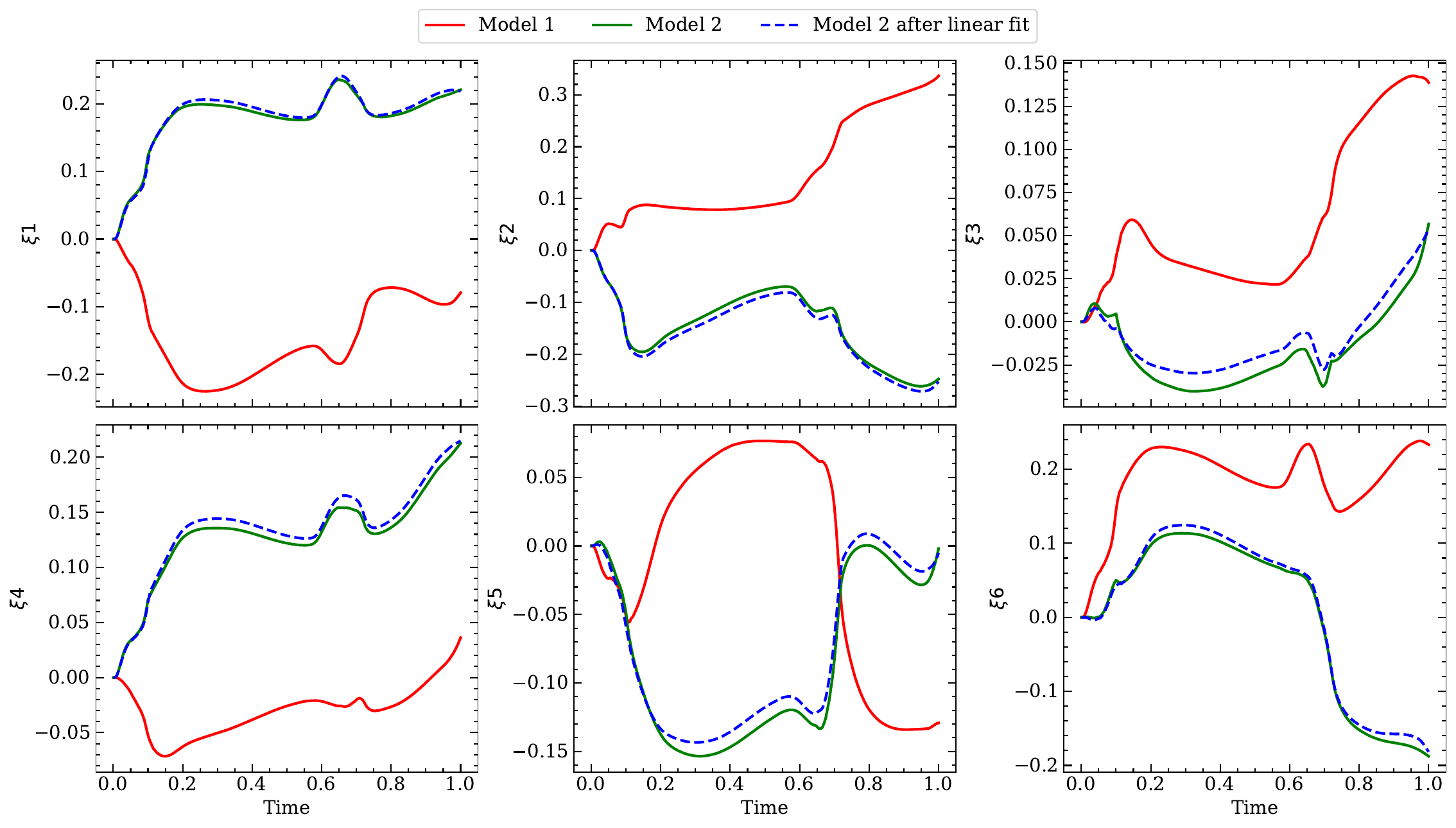}
    \caption{Visualization of the linear fit between two sets of internal variables.}
    \label{fig:iv_comparison}
\end{figure}

The identifiability of internal variables up to a linear transform is discussed.  As the internal variables are a physically unobservable quantity, there is nothing to compare the predicted internal variable against. However, as discussed in Section~\ref{sec:Int}, the predicted trajectory of internal variables, for a given trajectory of $\tensor{C}$, should be unique up to scaling. Therefore, we train two different NN models on the same dataset, with shared training and testing samples and compare the predicted internal variables by the two models. Figure~\ref{fig:iv_comparison} shows the visual comparison between the sets of internal variables predicted using two independently trained NN models, dissimilar in their initialization of parameters.  It is observed from Figure~\ref{fig:iv_comparison} that, although the two sets of internal variables (labelled as Model 1  and Model 2) do not match, they indeed are linear transformations of one another.  Moreover,  Figure~\ref{fig:iv_comparison} shows the quality of this linear fit using the relative error between two sets of internal variables. Note that the linear regression model is fitted on the internal variables corresponding to the training set, and then the same linear regression model is used to check if this scaling holds for the test set. The median error in the linear fit is approximately $3\%$ on both the train and test sets.  

\begin{figure}[H]
     \centering
     \includegraphics[width=\textwidth]{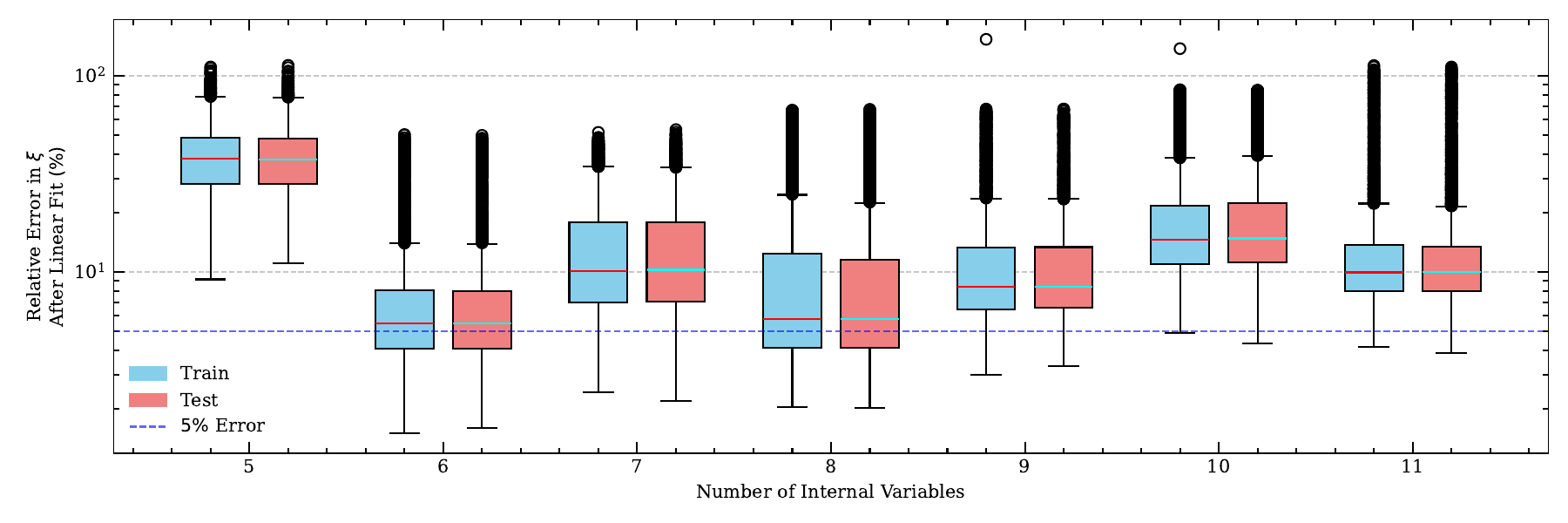}
        \caption{Error in the linear fit between two sets of internal variables obtained by independently training the same architecture with different initializations.}
        \label{fig:iv_orbit}
\end{figure}

\section{Conclusion}
\label{sec:conclusion}
In this work, we proposed an energy-based NN framework to model a history-dependent constitutive law.  The framework, which relies on the internal variable theory, was designed to be causal, interpretable, and incorporate knowledge from physics, namely the second law of thermodynamics and material stability, and consistency with the existence theory of the non-linear equilibrium problem.  We outlined methods for the NN-based constitutive law to ensure consistency with material stability and the second law of thermodynamics. We demonstrated the efficacy of this framework in homogenizing a polycrystalline material with a viscoelastic response. An accuracy of $2\%$  in terms of relative error was achieved in the prediction of stress trajectories. Moreover, we theoretically showed that the internal variables obtained as by-products of stress prediction are unique up to a scaling factor. This claim was empirically verified by fitting a linear regression model. 
While the method is deployed to model viscoelastic response, it can also capture any history-dependent response. These machine learning models can be directly used as a surrogate of the constitutive law of a unit cell in a multiscale FEM-based equilibrium equation solver such as Abaqus. Further research of interest includes:
(a) learning heterogeneous, microstructure-dependent, potentials; and (b) learning a spatially non-local response. 

\section*{Code and Data Availability}
The code is available at \url{https://github.com/m-raj/Learning-Stable-Material-Model.git}. Data will be made available upon request. 
\section*{Acknowledgments}

We gratefully acknowledge the financial support of the US Office of Naval Research through MURI grant  N00014-23-1-2654. The computations presented here were conducted in the Resnick High Performance Computing Center, a facility supported by Resnick Sustainability Institute at the California Institute of Technology.
\bibliographystyle{plainnat} 
\bibliography{references}
\begin{appendices}
\section{Proofs}

\begin{customthm}{\ref{th:existence_of_solutions}}
Given a pair of functions ($\cW$, $\widehat{\cD}$) satisfying Assumption~\ref{assump:w_d}. For any $F \in L^\infty$, consider  $S(t)$ and  $\xi(t)$ given by $\eqref{eq:sode}$. Then,  for every $A \in \mathrm{GL}(n, \R)$, there exist functions
\begin{subequations}
    \begin{alignat}{3}
        \widetilde{\cW}&:\R^m \times \R^n \to \R^+,\quad&(f, \widetilde{p}) &\mapsto \widetilde{\cW}(f, \widetilde{p});\\
        \widetilde{\cD} &: \R^n \to \R^+,\quad  &\widetilde{q} &\mapsto \widetilde{\cD}(\widetilde{q})
    \end{alignat}
\end{subequations}
    with $\widetilde{\cS}: \R^m \times \R^n \to \R^m$ and $\widetilde{\cF}: \R^m \times \R^n \times \R^n \to \R^n$, defined by,  
    \begin{subequations}
        \begin{align}
            \widetilde{{\cS}}(f, \widetilde{p}) &\coloneqq \dfrac{\partial \widetilde{\cW}(f, \widetilde{p})}{\partial c} 
\\
            \widetilde{\cF}(f, \widetilde{p}, \widetilde{q}) &\coloneqq \dfrac{\partial \widetilde{\cD}(\widetilde{q})}{\partial \widetilde{q}}+\dfrac{\partial \widetilde{\cW}(f, \widetilde{p})}{\partial \widetilde{p}}  
        \end{align}
    \end{subequations}
such that, for $\eta(t) := A\xi(t)$ for $t \in [0, T]$,
\begin{subequations}
        \begin{align}
    S(t) &= \widetilde{\cS}(F(t), \eta(t)); \\
    \widetilde{\cF}(F(t), \eta(t), \dot{\eta}(t)) &= 0,\quad \eta(0) = 0.
    \end{align} 
\end{subequations}
Moreover, the convexity of $\widehat{\cD}$ is preserved in $\widetilde{\cD}$.`
\end{customthm}
\begin{proof}
    We define 
    \begin{subequations}
    \begin{align}
        \widetilde{\cW}(f, \widetilde{p}) &\coloneqq \cW(f, p)\\
        \widetilde{\cD}(\widetilde{q}) &\coloneqq \widehat{\cD}(q)
    \end{align}
    \end{subequations}
such that 
\begin{align}
    \widetilde{p} = A p,\quad
    \widetilde{q} = Aq.
\end{align} Hence, 
\begin{align}
    \widetilde{\cS}(f, \widetilde{p}) &\coloneqq \dfrac{\partial \widetilde{\cW}(f, \widetilde{p})}{\partial c} = \dfrac{\partial \cW(f, p)}{\partial c} = \cS(f, p).
\end{align}
and 
\begin{align}
    A^{-\top}\dfrac{\partial \cW(f, p)}{\partial p} + A^{-\top}\dfrac{\partial\widehat{\cD}(q)}{\partial q} = A^{-\top} \cF(f, p, q)
\end{align}
Substituting $f = F(t)$, $p = \xi(t)$ and $q = \dot{\xi}(t)$, we get 
\begin{subequations}
\begin{align}
S(t) &= \cS(F(t), \xi(t)) = \widetilde{\cS}(F(t), A\xi(t))\\
\widetilde{\cF}(F(t), A\xi(t), A\dot{\xi}(t)) &= A^{-\top}\cF(F(t), \xi(t), \dot{\xi}(t)). 
\end{align}
\end{subequations}
We define $\eta(t) = A\xi(t)$, and use \eqref{eq:sodeb} to conclude, 
\begin{subequations}
    \begin{align}
        S(t) &= \dfrac{\widetilde{\cW}(F(t), \eta(t))}{\partial c}\\
        \widetilde{\cF}(F(t), \eta(t), \dot{\eta}(t)) &= \dfrac{\widetilde{\cW}(F(t), \eta(t))}{\partial \widetilde{p}} + \dfrac{\partial \widetilde{\cD}(\dot{\eta}(t))}{\partial \widetilde{q}} = 0, \quad \eta(0) = 0. \end{align}
\end{subequations}
\end{proof}
\begin{lemma}
\label{eq:diffeo}
    Let  $\cQ(p)$ represent an orthogonal matrix, parameterized by $p$; and  let $\cT(p)_{i,j}$ represent $\partial \cT(p)_i/\partial p_j $. If $\cT(p)_{i,j} = \cQ(p)_{ij}$,  then $\cT(p)_{ij} = [\cQ_0]_{ij}$ where $\cQ_0$ is an orthogonal matrix independent of $p$.
\begin{proof}
We have  given that $\cT(p)_{i,j} = \cQ(p)_{i,j}$.  Differentiating both sides with $p_k$, we have
\begin{align}
    \cT(p)_{i,jk} &= \cQ(p)_{ij,k} \label{eq:def}.
\end{align}
    By orthogonality of $\cQ(p)$, we have 
        \begin{align} 
        \cQ(p)_{ij}\cQ(p)_{jm} &= \delta_{im};\\
        \implies \cQ(p)_{ij,k} + \cQ(p)_{jm,k} &= 0\quad \forall i,m;\quad (\text{Differentiating both sides with }p_k),\\ 
        \implies \cQ(p)_{ij,k} &= - \cQ(p)_{ji,k}; \quad (m=i). \\
        \implies \cT(p)_{i,jk} &= - \cT(p)_{j,ik} \quad (\text{using } \eqref{eq:def}) \label{eq:anti-sym}
    \end{align}
As a mixed partial derivative of $\cT$ is continuous, we have 
\begin{align}
    \cT(p)_{i,jk} = \cT(p)_{i,kj}. \label{eq:mixed_partial}
\end{align}
Now, we used \eqref{eq:anti-sym} and \eqref{eq:mixed_partial}, 
\begin{subequations}
\begin{align}
    \cT(p)_{i,jk} &= \cT(p)_{i,kj}\\
    &= -\cT(p)_{k,ij}\\
    &= -\cT(p)_{k,ji}\\
    &= \cT(p)_{j,ki}\\
    &= \cT(p)_{j,ik}\\
    &= -\cT(p)_{i,jk}.
\end{align}
\end{subequations}
Hence, $\cT(p)_{i,jk} = - \cT(p)_{i,jk}$. Hence, $\cT(p)_{i,jk} \equiv 0.$ Hence, $\cT(p)_{i,j} = [Q_0]_{ij}$ where $Q_0$ is independent of $p$.
\end{proof}
\end{lemma}
\begin{proposition}
\label{prop:convex_fn}
    Let $P, Q$ be open sets in $\R^n$. Let $\cD$ and $\widetilde{\cD}$ be two strongly convex functions with minima at zero, and their minimum values be zero. Let $\cT : P \to \R^n$ be a diffeomorphism. If
    \begin{align}
        \cD(q) = \widetilde{\cD}(A(p)q), \quad \forall p, q \in P \times Q, \label{eq:convex_prop}
    \end{align}
    where $A(p) \coloneqq [\partial \cT(p)_i/\partial p_j]$; then 
    $\cT(p) = A_0 p$ is linear for $A_0 \in \mathrm{GL}(n, \R)$.
    \begin{proof}
As \eqref{eq:convex_prop} is true for all $p, q \in P \times Q$, we pick $p_1$ and $p_2$ such that $A(p_1) \neq A(p_2)$. Hence, we have
\begin{subequations}
\begin{align}
    \cD(q) &= \widetilde{\cD}(A(p_1)q), \quad \forall q \in Q; \label{eq:eq1}\\
    \cD(q) &= \widetilde{\cD}(A(p_2)q),\quad \forall q \in Q. \label{eq:eq2}
\end{align}
\end{subequations}

Using \eqref{eq:eq1} and \eqref{eq:eq2}, we have 
\begin{align}
    \widetilde{\cD}(A(p_1) q) &= \widetilde{\cD}(A(p_2)q),\quad \forall q \in Q.
\end{align}
As $A(p_1)$ is invertible, substitute $q = A(p_1)^{-1}\widetilde{q}$ to obtain 
\begin{align}
    \widetilde{\cD}(\widetilde{q}) = \widetilde{\cD}(A(p_2)A(p_1)^{-1}\widetilde{q}). \label{eq:identity}
\end{align}
Now, for $\alpha > 0$, we define the level set $\widetilde{Y}_\alpha$, as 
\begin{align}
 \widetilde{Y}_\alpha \coloneqq \{\widetilde{q} : \widetilde{\cD}(\widetilde{q}) = \alpha \}. \label{eq:level_sets}
\end{align}
Hence, \eqref{eq:identity} implies that 
\begin{align}
    \widetilde{q} \in \widetilde{Y}_\alpha \iff A(p_2)A(p_1)^{-1}\widetilde{q} \in \widetilde{Y}_\alpha
\end{align}
Hence, we have 
\begin{align}
    A(p_2)A(p_1)^{-1}: \widetilde{Y}_\alpha \to \widetilde{Y}_\alpha, \label{eq:bijection}
\end{align}
is a bijection. Observing that $\widetilde{Y}_\alpha$ is a bounded set, it follows that 
\begin{align}
    A(p_2)A(p_1)^{-1} = \cQ_\alpha(p) \quad \forall p_1, p_2 \in P,\quad \alpha \in \R^+. \label{eq:diff}
\end{align}
Hence, 
\begin{align}
    A(p) = \cQ_\alpha(p) A_0; \quad A_0 \in \mathrm{GL}(n, \R).
\end{align}
We apply Lemma~\ref{eq:diff} on the diffeomorphism $\cT \circ \cT_0$, where $\cT_0(p) \coloneqq A_0 p$, to deduce that $\cQ_\alpha(p)$ must be independent of $p$. Also, we denote $\widetilde{A}(\alpha) \coloneqq Q_\alpha A_0$, to get 
\begin{align}
    A(p) = \widetilde{A}(\alpha),\quad \forall p \in P, \forall \alpha \in \R^+. 
\end{align}
Hence, it must be that 
\begin{align}
    A(p) \equiv \widetilde{A}(\alpha) \equiv A_0.
\end{align}
where $A_0 \in \mathrm{GL}(n, \R)$. Hence $\cT$ is affine such that $\cT(p) = A_0 p + p_0$. But as $\cD$ and $\widetilde{\cD}$ have minima at zero, $p_0=0$. Hence, $\cT(p) = A_0 p$, is linear.
\end{proof}
\end{proposition} 
From the Example~\ref{ex:3}, it is observed that Assumption~\ref{assup:controllability} is justified for quadratic $\cW$ and convex $\cD$, when $m \geq n$. On the other hand, from example~\ref{ex:4}, the assumption is not justified when $n > m$, and $\cW$ is a quadratic function. In the example~\ref{ex:3}, the non-linearity comes from $\cD$. 
\begin{lemma}
    $U  = \R^m$.
    \begin{proof}
        For any $c_* \in \R^m$ and fixed $t=t_*$, we can construct $C_* \in \cC^1([0,T]; \R^m)$, given by $C_* (t) = f_*t/t_*$, such that $C_*(t_*)= f_*$, $C_*(0) =0$. 
    \end{proof}
\end{lemma}

\begin{customthm}{\ref{th:main_th}}
    For given two pairs of functions $(\cW, \widehat{\cD})$ and $(\widetilde{\cW}, \widetilde{\cD})$; consider ($\cS$,  $\cF$), and $(\widetilde{\cS}$, $\widetilde{\cF})$ defined as per \eqref{eq:seq}, and  \eqref{eq:stildeeq} respectively. Let for all $C \in L^\infty$ and $t \in [0, T]$,  
    \begin{subequations}
        \begin{align}
        \cS(F(t), \xi(t)) &= \widetilde{\cS}(F(t), \eta(t)) \label{eq:app_ode0}\noeqref{eq:app_ode0}, \\
        \cF(F(t), \xi(t), \dot{\xi}(t)) &= 0, \quad \xi(0) = 0, \label{eq:app_aode1}\\
        \widetilde{\cF}(F(t), \eta(t), \dot{\eta}(t)) &= 0, \quad \eta(0) = 0.\label{eq:app_aode2}
    \end{align}
    \end{subequations}
Moreover, let $\cF$ satisfy Assumption~\ref{assup:controllability} and $P \subset \R^n$ be the set as defined in Assumption~\ref{assup:controllability}. Let $\cT : P \to \R^n$ be a smooth diffeomorphism such that $\eta(t) = \cT(\xi(t))$. Then, $\cT$ must be  linear i.e. $\cT(\xi(t)) = A\xi(t)$ for all $\xi(t) \in P$, where $A \in \mathrm{GL}(n, \R)$. 
\begin{proof} 
Using Assumption~\ref{assup:controllability}, we restrict the functions $\cW$ and $\cD$ as follows
\begin{subequations}
\begin{align}
    \cW\vert_{U\times P} &: U\times P \to \R^+\\
    \widehat{\cD}\vert_Q&: Q \to \R^+.
\end{align}
\end{subequations}
We work with the restricted functions for the rest of the proof. For notational convenience, in this proof, we continue to use $\cW$ and $\cD$ instead of $\cW|_{U \times P}$ and $\cD|_{Q}$, respectively. 
    \begin{align}
            \cS(F(t), \xi(t)) &= \widetilde{\cS}(F(t), \eta(t))\\
            \implies \dfrac{\partial \cW (F(t), \xi(t))}{\partial c} &= \dfrac{\partial \widetilde{\cW}(F(t), \eta(t))}{\partial c}\\
            \text{Integrating on the set } U, &\\
            \implies \cW(F(t), \xi(t)) &= \widetilde{\cW}(F(t), \cT(\xi(t))) + h(\xi(t))
        \end{align}
        where $h:\R^m \to \R$ is constant of integration w.r.t, $f$. Further, differentiating both sides, 
        \begin{align}
            \dfrac{\partial \cW(F(t), \xi(t))}{\partial p} &= \nabla \cT(\xi(t))^{\top}\dfrac{\partial \widetilde{\cW}(F(t), \cT(\xi(t)))}{\partial \widetilde{p}} + \nabla h(\xi(t))
\end{align}
Using \eqref{eq:app_aode1}, \eqref{eq:app_aode2} and Assumption~\ref{assup:controllability}
\begin{align}
            \implies \dfrac{\partial \widehat{\cD}(q)}{\partial q}= \nabla k(p)^{\top}\dfrac{\partial \widetilde{\cD}(\nabla k(p) q)}{\partial \widetilde{q}} &+ \nabla h(p), \quad \forall p, q \in P\times Q \subset \R^n \times \R^n. 
        \end{align}
Upon substituting $q=0$ and using the fact that $\widehat{\cD}$ and $\widetilde{\cD}$ have minima at zero, we conclude $\nabla h(p) = 0,\quad \forall p$. Hence, $h(p) = h_0$ is a constant function. Hence, we have 
\begin{align}
    \dfrac{\partial \widehat{\cD}(q)}{\partial q} &= \nabla k(p)^{\top}\dfrac{\partial \widetilde{\cD}(\nabla k(p) q)}{\partial \widetilde{q}},\label{eq:dd}
    \end{align}
  Integrating on the set $Q$
    \begin{align}\implies \widehat{\cD}(q) &= \widetilde{\cD}(\nabla k(p) q) + \widetilde{h}(p).
\end{align}
where $\widetilde{h}(p)$ is constant of integration with respect to $p$. Using that minima of $\widehat{\cD}$ and $\widetilde{\cD}$ are zero, upon substituting $q=0$, we conclude $\widetilde{h}(p) \equiv 0$. Hence, 
\begin{align}
    \widehat{\cD}(q) = \widetilde{\cD}(\nabla k(p)q), \quad \forall p, q \in P \times Q. 
\end{align}
Using Proposition~\ref{prop:convex_fn}, we conclude $\nabla \cT(p) \equiv A_0.$ 
    \end{proof}
\end{customthm}
\end{appendices}
\end{document}